\documentclass[%
reprint,  
 amsmath,amssymb,
 aps,
]{revtex4-2}

\usepackage[dvipdfmx]{graphicx}
\usepackage{dcolumn}
\usepackage{bm}
\usepackage{color}
\usepackage{siunitx}

\usepackage{physics}

\begin{document}

\preprint{APS/123-QED}

\title{Pair dynamics of active force dipoles in an odd-viscous fluid}

\author{Yuto Hosaka}\email{yuto.hosaka@ds.mpg.de}
\affiliation{Max Planck Institute for Dynamics and Self-Organization (MPI DS), Am Fa{\ss}berg 17, 37077 G\"{o}ttingen, Germany}

\author{David Andelman}\email{andelman@tauex.tau.ac.il}
\affiliation{School of Physics and Astronomy, Tel Aviv University, Ramat Aviv, Tel Aviv 69978, Israel}

\author{Shigeyuki Komura}\email{komura@wiucas.ac.cn}
\affiliation{Wenzhou Institute, University of Chinese Academy of Sciences, Wenzhou, Zhejiang 325001, China}
\affiliation{Oujiang Laboratory, Wenzhou, Zhejiang 325000, China}
\affiliation{Department of Chemistry, Graduate School of Science, Tokyo Metropolitan University, Tokyo 192-0397, Japan}


\begin{abstract}
We discuss the lateral dynamics of two active force dipoles, which interact with each other via hydrodynamic interactions in a thin fluid layer that is active and chiral.
The fluid layer is modeled as a two-dimensional (2D) compressible fluid with an odd viscosity, while the force dipole (representing an active protein or enzyme) induces a dipolar flow.
Taking into account the momentum decay in the 2D fluid, we obtain analytically the mobility tensor that depends on the odd viscosity and includes nonreciprocal hydrodynamic interactions.
We find that the particle pair shows spiral behavior due to the transverse flow induced by the odd viscosity.
When the magnitude of the odd viscosity is large as compared with the shear viscosity, two types of oscillatory behaviors are seen.
One of them can be understood as arising from closed orbits in dynamical systems, and its circular trajectories are determined by the ratio between the magnitude of the odd viscosity and the force dipole.
In addition, the phase diagrams of the particle dipolar angles are obtained numerically.
Our findings reveal that the nonreciprocal response leads to complex dynamics of active particles embedded in an active fluid with odd viscosity.
\end{abstract}

\maketitle


\section{Introduction}

Enzymes are nanometer-size biomolecular complexes that catalyze biochemical reactions and play a crucial role in various life-sustaining processes.
For example, they facilitate molecular transport into the cell and assist chemical reactions that are essential for cellular metabolism and homeostasis~\cite{albertsbook}.
In recent years, nonequilibrium transport phenomena induced by enzymes have attracted a considerable attention due to novel applications such as drug delivery and the design of synthetic nanomachines~\cite{gompper2020}.

In aqueous solutions of enzymes, enhanced diffusion was experimentally reported for certain enzymes~\cite{riedel2015,illien2017}, as well as for passive probe particles~\cite{zhao2017} in the presence of \textit{substrate} molecules.
Here substrates are chemical species that react with enzymes and are converted into product molecules.
When a spacial gradient in the substrate molecule concentration exists, enzymes exhibit a collective motion in the direction of higher or lower concentrations.
This phenomenon is known as chemotaxis~\cite{sengupta2013,sengupta2014,yu2009,dey2014} or antichemotaxis, respectively~\cite{jee2018}.
Although a molecular diffusion enhancement was observed also for organic chemical reactions at much smaller length scales (subnanometer)~\cite{dey2016,wang2020}, the observed diffusion enhancement is still a matter of debate~\cite{macdonald2019,rezaei2022} and more experiments are needed.

To better understand enzyme mobility, various theoretical studies using coarse-grained models have been developed.
In previous works, an enzyme was modeled as built of two sub-units representing its overall structure.
The two units are connected by an elastic spring reflecting inherent enzyme relaxation dynamics~\cite{togashi2007}.
By using equilibrium approaches, it was demonstrated that the internal degrees of freedom leads to enhanced diffusion of the enzyme in solution~\cite{illien2017_epl,hosaka2020}.
On the other hand, in nonequilibrium situations, enzyme conformational dynamics 
collectively induces hydrodynamic flows~\cite{hosaka2020_2,hosaka2017,mikhailov2015}.
Furthermore, employing a force dipole model for the enzyme, it was found that such nonequilibrium effects lead to an increase in particle diffusion in solutions as well as for biological membranes~\cite{hosaka2020_2,hosaka2017,mikhailov2015}.
More recently, hydrodynamic interactions between force dipoles were taken into account, and clustering mechanisms of force dipoles were investigated both in flat~\cite{manikantan2020} or curved~\cite{bagaria2022} membrane geometries.

These enzymatic nonequilibrium effects were studied only in passive fluids, whereas at physiological conditions, other sources of nonequilibrium effects can turn the surrounding fluids into active fluids.
For example, membrane proteins such as ATPase can autonomously rotate in the presence of ATP (adenosine triphosphate) or proton gradients, and their induced hydrodynamic flows drive the surrounding membrane into an out-of-equilibrium state~\cite{oppenheimer2019,manneville1999,manneville2001}.
More specifically, ATP consumption and autonomous rotation of membrane proteins can lead to breaking of time-reversal and parity symmetries in membranes~\cite{banerjee2017}.
These ATP consumption and autonomous rotation endow membranes with active and chiral features, respectively.
At length scales larger than the mean distance between rotary proteins, membranes with these proteins can be viewed as two-dimensional (2D) active chiral fluids.
In such active chiral 2D systems, it is known that a dissipationless transport coefficient called the \textit{odd viscosity} emerges~\cite{hosaka2022,fruchart2022}.
To reveal the hydrodynamic effects of odd viscosity, its consequence has been studied for the motion of passive objects~\cite{ganeshan2017,souslov2020,hosaka2021,hosaka2021_2,lier2022}, many-body sedimentation~\cite{khain2022}, and density waves~\cite{banerjee2017,markovich2021}.
One of the peculiar features of active chiral fluids is their nonreciprocal interaction~\cite{hosaka2021,hosaka2021_2,khain2022,lier2022} that is prohibited for passive fluids.
However, despite these intriguing findings, the dynamics of active enzymes in an active chiral environment was not studied, and the role of odd viscosity in biomembranes remains unexplored.

In this paper, we discuss the lateral dynamics of an enzyme pair that interacts via hydrodynamic interactions as it is embedded in a thin layer of an active chiral fluid.
The enzymes are modeled as active particles that induce force dipoles~\cite{mikhailov2015,manikantan2020}.
To investigate collective behavior of \textit{active} particles in an \textit{active} environment, we consider force dipoles in a chiral fluid layer that is modeled as a 2D compressible fluid with odd viscosity.
Since such fluids have a nonreciprocal nature, active particles are expected to exhibit chiral trajectories although the particles themselves are apolar and do no have any preferred direction.

Extending our previous work~\cite{hosaka2021}, we derive the odd viscosity-dependent mobility tensor that includes nonreciprocal hydrodynamic interactions.
As a minimum model to explore hydrodynamic interactions between active particles, we consider pairs of two active particles and discuss the nonlinear two-body dynamics.
In contrast to the passive fluid case without odd viscosity, we find that active particle pairs show spiral trajectories where one particle follows the other.
When the magnitude of the odd viscosity is large as compared with the shear viscosity, particles show two types of periodic oscillations, including one which is determined by the ratio between the magnitude of the odd viscosity and force dipole.
Our findings reveal that the nonreciprocal response due to odd viscosity leads to complex dynamics of active particles.

The outline of the manuscript is as follows.
In Sec.~\ref{sec:system}, we introduce the hydrodynamic equations for a 2D active chiral fluid and derive its mobility tensor~\cite{hosaka2021}.
In Sec.~\ref{sec:twobody}, we obtain the nonlinear equations for the distance between the two active particles and their relative polar angles~\cite{manikantan2020}.
Using these nonlinear equations, we analyze in Sec.~\ref{sec:result} the dynamics of particle pairs for various values of odd viscosity and classify their behavior into several characteristic states.
Section~\ref{sec:discussion} includes further discussions, and a summary and conclusions are presented in Sec.~\ref{sec:conc}.

\begin{figure}[htb]
\centering
\includegraphics[scale=.3]{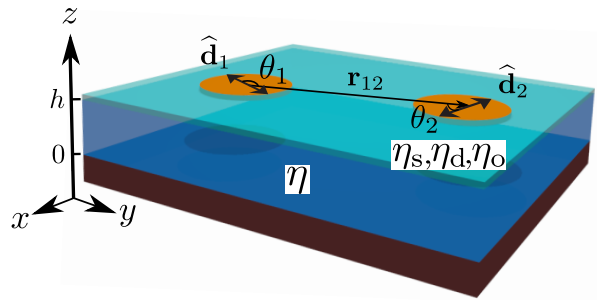}
\caption{
Schematic sketch of an active and chiral 2D fluid layer.
The infinitely large, flat, and thin 2D fluid layer (light blue) is located at $z=h$ and has 2D shear, dilatational, and odd viscosities, $\eta_{\rm s}$, $\eta_{\rm d}$, and $\eta_{\rm o}$, respectively.
This fluid layer is in contact with air $(z>h)$ and a 3D fluid (dark blue) underneath $(0<z<h)$ characterized by a 3D shear viscosity $\eta$.
The 3D fluid is bounded from below by an impermeable flat solid surface (brown) located at $z=0$, and the 3D velocity is assumed to vanish at $z=0$.
The active particle that represents an enzyme (orange disk) has radius $a$ and can move laterally within the 2D fluid layer.
The dipolar flow that is induced by particle ``1'' (``2'') is characterized by its direction $\widehat{\mathbf{d}}_1$ $(\widehat{\mathbf{d}}_2)$ and the relative angle $\theta_1$ $(\theta_2)$ as defined in Eq.~(\ref{eq:cos}).
}
\label{fig:system}
\end{figure}

\section{Active chiral 2D layer}
\label{sec:system}

\subsection{Hydrodynamic equations}

We consider an active, chiral, and compressible 2D layer, which is flat, very thin, and infinitely large, as schematically depicted in Fig.~\ref{fig:system}.
The layer is in contact with a 3D incompressible fluid (e.g., water), which is bound from below by a rigid substrate. 
This leads to a momentum leakage from the 2D layer to the 3D fluid.
We assume that the 3D fluid thickness $h$ is much smaller than any in-plane characteristic length scales so that 
the lubrication approximation will hold for the 3D fluid~\cite{barentin1999,elfring2016}.
The 2D compressible fluid layer can be realized experimentally, for example, by a dilute Gibbs monolayer composed of soluble amphiphiles that can dissolve into the underlying 3D fluid~\cite{barentin1999}.
Such a 2D/3D material transport makes the 2D fluid compressible.
We assume that the adsorption and desorption processes of soluble amphiphiles are instantaneous, and can be regarded as a limiting case of a finite relaxation time~\cite{lier2022}.

The active chiral nature of the 2D fluid is taken into account by introducing the concept of odd viscosity.
Although the regular viscosity (such as shear and dilatational viscosities) is always positive, odd viscosity can be either positive or negative, depending on the chirality direction.
At microscopic scales, the 2D odd viscosity sign is related to the rotational direction of the active constituents such as rotary proteins in biological membranes~\cite{markovich2021,banerjee2017}.
To see clearly the effect of odd viscosity on the active particle dynamics, we consider a compressible 2D fluid~\cite{hosaka2021}.
The reason being that the odd viscosity does not play a role in the velocity field of an incompressible fluid~\cite{ganeshan2017}.
Note that in the incompressible layer limit, the fluid flow becomes independent of the odd viscosity, as will be shown below.

We denote the 2D velocity field by $\mathbf{v}(\mathbf{r})$ with $\mathbf{r}=(x,y)$ being a positional vector in 2D, and $p$ is the hydrostatic pressure of the 3D fluid.
At low Reynolds numbers, the momentum balance equation for the 2D fluid can be written as~\cite{hosaka2021}
\begin{align}
\eta_{\rm s}\nabla^2\mathbf{v} +\eta_{\rm d}\nabla (\nabla\cdot\mathbf{v}) + \eta_{\rm o}\nabla^2 \mathbf{v}^\ast 
 -\frac{h}{2}\nabla p
+\mathbf{f}^{\rm 3D}
+\mathbf{F}=0.
\label{eq:stokes}
\end{align}
Here, $\eta_{\rm s}$, $\eta_{\rm d}$, and $\eta_{\rm o}$ are the 2D shear, dilatational, and odd viscosities, respectively, $\nabla=(\partial_x, \partial_y)$ stands for the 2D gradient operator, $\mathbf{f}^{\rm 3D}$ is the vectorial force density exerted on the 2D fluid layer by the underlying 3D fluid, $\mathbf{F}$ is any other force density acting on the 2D fluid, $v_i^\ast=\epsilon_{ij}v_j$ is the velocity vector rotated clockwise by $\pi/2$, and $\epsilon_{ij}$ is the 2D Levi-Civita tensor with $\epsilon_{xx}=\epsilon_{yy}=0$ and $\epsilon_{xy}=-\epsilon_{yx}=1$.
Notice that the force density $\mathbf{f}^{\rm 3D}$ is given by the projection of the 3D fluid traction on the $xy$ plane.
Thus, $\mathbf{f}^{\rm 3D}$ is a 2D vector parallel to the 2D layer.
In general, $\mathbf{f}^{\rm 3D}$ can be expressed as $f_i^{\rm 3D}=-(\zeta_\|\delta_{ij}+\zeta_\perp\epsilon_{ij})v_j$~\cite{soni2019}, where $\zeta_\|$ is the friction coefficient representing the momentum dissipation parallel to $\mathbf{v}$, while $\zeta_\perp$ acts perpendicular to $\mathbf{v}$, and $\delta_{ij}$ is the Kronecker delta.

The divergence of the in-plane velocity is given by~\cite{barentin1999,elfring2016}
\begin{align}
\nabla\cdot\mathbf{v} = \frac{h^2}{6\eta}\nabla^2 p,
\label{eq:compress}
\end{align}
where $\eta$ is the shear viscosity of the underlying 3D fluid.
Equation~(\ref{eq:compress}) can be derived by taking the divergence of the underlying 3D fluid velocity and 
integrating over its thickness $(0\leq z\leq h)$~\cite{barentin1999}.
This derivation relies on the lubrication approximation that is justified when the 3D fluid is shallow enough 
so that the vertical component of the 3D velocity can be neglected as compared to its in-plane components.
When the thickness $h$ is finite, however, the 2D and 3D Stokes equations are coupled
to each other~\cite{manikantan2020surfactant} and a numerical treatment is required~\cite{stone1998}.
For analytical tractability, we do not consider such an intermediate situation in this work.

\subsection{Mobility tensor of an active chiral layer}

The force density $\mathbf{F}$ acting on the fluid layer at position $\mathbf{r}^\prime$ is connected via a second-rank mobility tensor $\mathbf{G}(\mathbf{r})$ with the induced fluid velocity at position $\mathbf{r}$:
\begin{align}
v_{i}(\mathbf{r})=\int {\rm d}^{2} r^{\prime}\, G_{i j}\left(\mathbf{r}-\mathbf{r}^{\prime}\right) F_{j}\left(\mathbf{r}^{\prime}\right).
\end{align}
Solving the coupled hydrodynamic equations~(\ref{eq:stokes}) and (\ref{eq:compress}) in Fourier space, we show in Appendix~\ref{app:GFT} that the mobility tensor $G_{ij}[\mathbf{k}]$ can be obtained as  
\begin{align}
&G_{ij}[\mathbf{k}] =\nonumber\\
&\frac{
\eta_{\rm s}(k^2+\kappa^2)\widehat{k}_i\widehat{k}_j
+(\eta_{\rm s}+\eta_{\rm d})(k^2+\lambda^2)\overline{k}_i\overline{k}_j
-\eta_{\rm o}(k^2+\nu^2)\epsilon_{i j}
}
{\eta_{\rm s}(\eta_{\rm s}+\eta_{\rm d})\left(k^{2}+\kappa^{2}\right)\left(k^{2}+\lambda^{2}\right)+\eta_{\rm o}^2\left(k^{2}+\nu^2\right)^2},
\label{eq:gmobilityk}
\end{align}
where $\mathbf{k} = (k_x ,k_y)$ is the 2D wavevector, $k=|\mathbf{k}|$, $\widehat{k}_i=k_i/k$, and $\overline{k}_i=-\epsilon_{ij}\widehat{k}_j$.
In the above, we have introduced three hydrodynamic screening lengths 
\begin{align}
\kappa^{-1}=\sqrt{\frac{\eta_{\rm s}}{\zeta_\|}},
\quad
\lambda^{-1}=\sqrt{\frac{h(\eta_{\rm s}+\eta_{\rm d})}{3\eta+h\zeta_\|}},
\quad
\nu^{-1}=\sqrt{\frac{\eta_{\rm o}}{\zeta_\perp}}.
\label{eq:screening}
\end{align}
Due to the transverse momentum decay $\zeta_\perp$, the screening length $\nu^{-1}$ depends only on the odd 
viscosity $\eta_{\rm o}$.
This screening length was not considered in our previous work~\cite{hosaka2021}.

In the special 2D incompressible limit $\eta_{\rm d}\to\infty$, Eq.~(\ref{eq:gmobilityk}) reduces to the mobility tensor for an incompressible supported 2D fluid~\cite{ramachandran2011,oppenheimer2010}
\begin{align}
G_{ij}^0[\mathbf{k}] =
\frac{\delta_{ij}-\widehat{k}_i\widehat{k}_j}
{\eta_{\rm s}\left(k^{2}+\kappa^{2}\right)}.
\end{align}
Note that $\mathbf{G}^0$ in this limit does not depend on the odd viscosity $\eta_{\rm o}$~\cite{ganeshan2017,khain2022,hosaka2021,souslov2020}.

For a 2D fluid layer supported by a rigid substrate, the parallel friction coefficient is given by $\zeta_\|\simeq
\eta/h$~\cite{evans1988} when $h$ is small enough as compared to all the three screening lengths in Eq.~(\ref{eq:screening}).
In order to obtain the real-space mobility tensor analytically, we assume that the three hydrodynamic screening 
lengths in Eq.~(\ref{eq:screening}) are all identical.
By setting $\eta_{\rm d}=3\eta_{\rm s}$ and assuming that the transverse friction $\zeta_\perp$ is proportional 
to the ratio between the odd and shear viscosities, i.e., $\zeta_\perp=(\eta_{\rm o}/\eta_{\rm s})\zeta_\|$, we 
obtain $\kappa^{-1}=\lambda^{-1}=\nu^{-1}=\eta_{\rm s}h/\eta$.
Then, Eq.~(\ref{eq:gmobilityk}) can be simplified as 
\begin{align}
G_{ij}[\mathbf{k}] =\frac{4\delta_{ij}-3\widehat{k}_i\widehat{k}_j- \mu\epsilon_{i j}}{\eta_{\rm s}(4+\mu^2)\left(k^{2}+\kappa^{2}\right)},
\label{eq:mobilityk}
\end{align}
where $\mu=\eta_{\rm o}/\eta_{\rm s}$ and the relation $\overline{k}_i\overline{k}_j=\delta_{ij}-\widehat{k}_i\widehat{k}_j$ has been used.
The dimensionless parameter $\mu$ is a measure of how far the 2D active chiral fluid departs from its passive analog, due to the active constituents that self-spin at microscopic scales.

The real-space representation of the mobility tensor can be obtained by the inverse Fourier transform of Eq.~(\ref{eq:mobilityk}).
The derivation is shown in Appendix~\ref{app:G} and the final result is
\begin{align}
G_{i j}(\mathbf{r})=C_{1}(r) \delta_{i j}+C_{2}(r) \widehat{r}_{i} \widehat{r}_{j}+C_{3}(r) \epsilon_{i j},
\label{eq:G}
\end{align}
where $\widehat{\mathbf{r}}=\mathbf{r}/r$ is a unit vector $(r=|\mathbf{r}|)$ and the three position-dependent coefficients are
\begin{align}
C_1(r) &= \frac{1}{2\pi\eta_{\rm s}(4+\mu^2)} \left[-\frac{3}{(\kappa r)^2}+4K_0(\kappa r)+\frac{3K_1(\kappa r)}{\kappa r}\right],
\nonumber\\
C_2(r) &= \frac{3}{2\pi\eta_{\rm s}(4+\mu^2)} \left[\frac{2}{(\kappa r)^2}-K_0(\kappa r)-\frac{2K_1(\kappa r)}{\kappa r}\right],
\label{eq:c123}\\
C_3(r) &=  -\frac{\mu K_0(\kappa r)}{2\pi\eta_{\rm s}(4+\mu^2)}
,\nonumber
\end{align}
and $K_n(x)$ is the modified Bessel function of the second kind~\cite{abramowitzhandbook}.
Note that $C_3$ exists only when $\mu\neq0$ (nonzero odd viscosity).

Expanding the mobility tensor, Eq.~(\ref{eq:G}), for $\kappa r\ll1$, we obtain
\begin{align}
G_{i j}(\mathbf{r}) &\approx \frac{1}{8\pi\eta_{\rm s}(4+\mu^2)}
\left[
\left(-3-10\gamma+ 10\ln\frac{2}{\kappa r}\right)\delta_{ij}\right.\nonumber\\
&\left.+6 \widehat{r}_{i} \widehat{r}_{j}
+4\mu\left(\gamma - \ln\frac{2}{\kappa r}\right)\epsilon_{ij}
\right].
\label{eq:Gsmall}
\end{align}
In this limit, the mobility tensor depends logarithmically on the distance $r$.
In the opposite limit of $\kappa r\gg1$, we obtain
\begin{align}
&G_{i j}(\mathbf{r})\approx  \frac{1}{2\pi\eta_{\rm s}(4+\mu^2)} \nonumber\\
&\times\left[
\frac{3}{(\kappa r)^2} \left(-\delta_{ij}+2\widehat{r}_{i} \widehat{r}_{j}\right)
-\mu
\sqrt{\frac{\pi}{2\kappa r}}e^{-\kappa r}\epsilon_{ij}
\right],
\label{eq:Glarge}
\end{align}
where $\gamma\approx0.5772$ is Euler's constant.
The term due to the regular viscosities in $\mathbf{G}$ decays algebraically ${\sim}(\kappa r)^{-2}$, while the term due to odd viscosity decreases exponentially ${\sim}e^{-\kappa r}$.

In the following, we concentrate on the regime $\kappa r\ll1$ in order to investigate the effect of odd viscosity on the collective dynamics of active particles.
The $\kappa r\ll1$ limit is justified when the product of the 2D shear viscosity $\eta_{\rm s}$ and the 3D fluid thickness $h$ is much larger than the 3D viscosity $\eta$, as occurring in physiological conditions.
Using typical values such as $\eta_{\rm s}\approx\SI[parse-numbers=false]{10^{-9}}{\pascal\second\metre}$, $h\approx\SI[parse-numbers=false]{10^{-8}}{\metre}$, and $\eta\approx\SI[parse-numbers=false]{10^{-3}}{\pascal\second}$~\cite{saffman1976,ramachandran2010}, we find $\kappa^{-1}\approx\SI[parse-numbers=false]{10^{-7}}{\metre}$.
When active particles move across length scales $\SI{1}{\nano\metre}\leq r\leq\SI{10}{\nano\metre}$, we have $10^{-2}\leq\kappa r\leq10^{-1}$ and the condition $\kappa r\ll1$ is satisfied.

\section{Two-body hydrodynamic interactions}
\label{sec:twobody}

\subsection{Hydrodynamic force dipole}

\begin{figure}[htb]
\centering
\includegraphics[scale=.7]{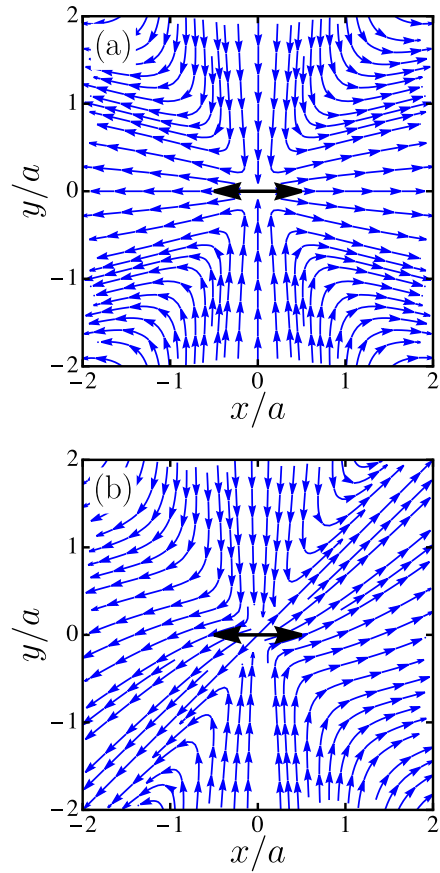}
\caption{
Streamlines of the 2D velocity $\mathbf{v}(x,y)$ generated by a force dipole.
The force dipole along the $x$-axis [$\widehat{\mathbf{d}}=(1,0)$] is centered at the origin (the black double arrow) for (a) $\mu=0$ (no odd viscosity, $\eta_{\rm o}=0)$ and (b) $\mu=1$ [see Eq.~(\ref{eq:velocity})].
The blue arrows indicate the flow direction.
}
\label{fig:dipole}
\end{figure}

In the 2D active chiral compressible fluid introduced above, we consider two active particles of radius $a$ that interact hydrodynamically with each other, as depicted in Fig.~\ref{fig:system}.
A single active particle can represent an enzymatic molecule.
Since enzymes generate dipolar flows while changing their conformations~\cite{mikhailov2015}, they can be modeled as active particles that induce hydrodynamic force dipole~\cite{manikantan2020,hosaka2017}.
Without any hydrodynamic interactions between other active particles, a single particle with a force dipole does not show any motility because the force-free condition is imposed on each particle.

When an active particle with a force dipole resides at the origin and its dipole is directed along a given unit vector $\widehat{\mathbf{d}}$, it induces a velocity field given by~\cite{mikhailov2015}
\begin{align}
v_{i}(\mathbf{r})=-\sigma \widehat{d}_{k} \partial_{k} G_{i j}(\mathbf{r}) \widehat{d}_{j}.
\label{eq:dipole}
\end{align}
Here, $\sigma=fa$ is the magnitude of the force dipole where $f$ is the force magnitude and $a\ll r$ is the distance between the two point forces (the particle size).

\begin{figure*}[htb]
\centering
\includegraphics[scale=.45]{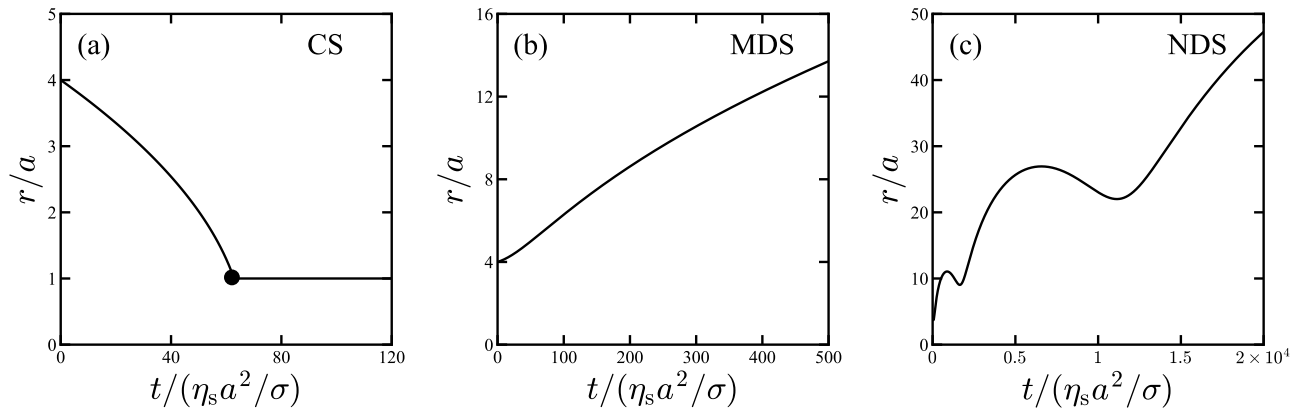}
\caption{
Plots of the rescaled distance between the two particles, $r=|\mathbf{r}_{12}|$, as a function of the rescaled time $t$ for $\eta_{\rm o}=0$.
(a) Convergence state (CS, cross in Fig.~\ref{fig:phase}) for $(\theta_1,\theta_2)=(\pi/2,\pi/2)$.
The black circle corresponds to the collision time between the two particles.
(b) Monotonic divergence state (MDS, closed triangle in Fig.~\ref{fig:phase}) for $(\theta_1,\theta_2)=(\pi/4,\pi/4)$.
(c) Nonmonotonic divergence state (NDS, open triangle in Fig.~\ref{fig:phase}) for $(\theta_1,\theta_2)=(\pi/2,\pi/4)$.
}
\label{fig:mu0}
\end{figure*}

Returning to the particle pair, the vector connecting particle ``1'' to particle ``2'' is denoted as $\mathbf{r}_{12}$, and the two relative angles $\theta_1,\theta_2$ are defined as 
\begin{align}
\cos\theta_1 = \frac{\widehat{\mathbf{d}}_1\cdot\mathbf{r}_{12}}{|\mathbf{r}_{12}|},~~~~~
\cos\theta_2 = \frac{\widehat{\mathbf{d}}_2\cdot\mathbf{r}_{21}}{|\mathbf{r}_{21}|},
\label{eq:cos}
\end{align}
where $\widehat{\mathbf{d}}_1$ $(\widehat{\mathbf{d}}_2)$ is the dipolar direction of particle $1$ $(2)$.
Substituting Eq.~(\ref{eq:Gsmall}) into Eq.~(\ref{eq:dipole}), we obtain the translational velocities of particle $2$ relative to particle $1$
\begin{align}
&\mathbf{v}_{21}(\mathbf{r})= \frac{\sigma_1}{4\pi\eta_{\rm s}(4+\mu^2)r}
\nonumber\\
&\times\left[
3\cos(2\theta_1)\widehat{\mathbf{r}}_{12}+2\cos\theta_1\widehat{\mathbf{d}}_1-2\mu\cos\theta_1\widehat{\mathbf{d}}_1^\ast
\right],
\label{eq:velocity}
\end{align}
with $\widehat{d}_i^\ast=\epsilon_{ij} \widehat{d}_j$. 
The derivation of Eq.~(\ref{eq:velocity}) is shown in Appendix~\ref{app:relative}.

The velocity field that is induced by a force dipole is plotted in Figs.~\ref{fig:dipole}(a) and (b) for $\mu=0$ $(\eta_{\rm o}=0)$ and $\mu=1$, respectively.
Here, the force dipole is located at the origin and is directed along the $x$-direction with $\widehat{\mathbf{d}}_1=(1,0)$.
When $\mu=0$ [as in Fig.~\ref{fig:dipole}(a)], the flow is symmetric with respect to both the $x$- and $y$- axes, and the azimuthal component $\widehat{\mathbf{d}}_1$ causes surrounding fluids to flow away from the origin.
This leads to an outward flow around the force dipole.
On the other hand, when $\mu=1$ [as in Fig.~\ref{fig:dipole}(b)], one sees that the emerging flow lines are tilted along the diagonal line $x=y$.
This symmetry breaking in the $x$- and $y$- axes is due to the perpendicular contribution $\widehat{\mathbf{d}}^\ast_1$ in Eq.~(\ref{eq:velocity}) when $\mu\neq0$.

\subsection{Pair dynamics of active particles}

We proceed by examining the hydrodynamic interactions between a pair of active particles, as depicted in Fig.~\ref{fig:system}.
Each particle moves laterally and rotates in the 2D fluid as a result of the fluid velocity induced by the other particle.
The separation between the two particles evolves as $\dot{r}=dr/dt=(\mathbf{v}_{21}-\mathbf{v}_{12})\cdot\widehat{\mathbf{r}}$~\cite{manikantan2020}, where $\mathbf{v}_{21}$ and $\mathbf{v}_{12}$ are the translational velocities of particle $2$ relative to $1$ [see Eq.~(\ref{eq:velocity})] and vice versa, and $\mathbf{r}=\mathbf{r}_{12}$.
When $\sigma_1=\sigma_2=\sigma$, the distance between the two particles evolves as
\begin{align}
&\dot{r}= \frac{\sigma}{4\pi\eta_{\rm s}(4+\mu^2)r} \nonumber\\
&\times
\biggl[2+4\cos(2\theta_1)+4\cos(2\theta_2)
-\mu\left[\sin(2\theta_1)+\sin(2\theta_2)\right] \biggl].
\label{eq:r}
\end{align}

Each particle rotates at a rate equal to half the vorticity of the flow velocity induced by the other particle: $\Omega_1=(\nabla\times\mathbf{v}_{12})_z/2$ and $\Omega_2=(\nabla\times\mathbf{v}_{21})_z/2$~\cite{manikantan2020}, where $\Omega_1$ and $\Omega_2$ are the $z$-components of vortices for the 2D rotation.
An additional effect of rotating $\mathbf{r}$ emerges when the induced velocity has an azimuthal component~\cite{manikantan2020}.
In total, the two angles evolve as
\begin{align}
\begin{aligned}
\dot{\theta}_1 &=-\frac{\sigma}{4\pi\eta_{\rm s}(4+\mu^2)r^2}[4\sin(2\theta_2)+\mu\cos(2\theta_2)]+\dot{\phi},
\\
\dot{\theta}_2 &=-\frac{\sigma}{4\pi\eta_{\rm s}(4+\mu^2)r^2}[4\sin(2\theta_1)+\mu\cos(2\theta_1)]+\dot{\phi},
\label{eq:theta12}
\end{aligned}
\end{align}
where $\dot{\phi}$ is an angular rotation due to the azimuthal components $\widehat{\mathbf{d}}$ and $\widehat{\mathbf{d}}^\ast$ of the velocity in Eq.~(\ref{eq:velocity}):
\begin{align}
&\dot{\phi}= -\frac{\sigma}{4\pi\eta_{\rm s}(4+\mu^2)r^2}\nonumber\\
&\times
\biggl[ \sin(2\theta_1)+\sin(2\theta_2)+\mu \left[2+\cos(2\theta_1)+\cos(2\theta_2)\right] \biggl].
\label{eq:phi}
\end{align}
More details about the derivation of Eq.~(\ref{eq:theta12}) are given in Appendix~\ref{app:relative}.
From Eqs.~(\ref{eq:theta12}) and (\ref{eq:phi}), one can see that the angular evolution is determined not only by the angle of the other particle, but also by its own angle because there is a coupling between $\theta_1$ and $\theta_2$, Eq.~(\ref{eq:phi}).
The asymptotic expressions of Eqs.~(\ref{eq:r}) and (\ref{eq:theta12}) for the two limits of $\mu\ll1$ and $\mu\gg1$ are also given in Appendix~\ref{app:relative}.

\begin{figure*}[htb]
\centering
\includegraphics[scale=.55]{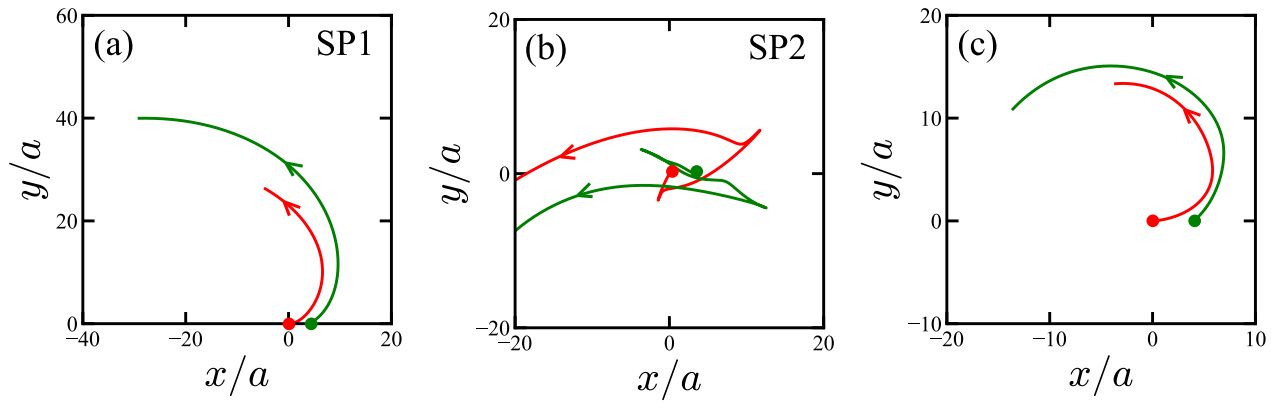}
\caption{
Pair dynamics of two active particles (denoted as red and green) initially located at $(x, y)=(0,0)$ and $(4a,0)$.
The arrows indicate the trajectory direction.
(a) Trajectories of a particle pair whose initial orientation is $(\theta_1,\theta_2)=(3\pi/4, \pi/2)$ for $\mu=3$ (SP1, orange diamond in Fig.~\ref{fig:phase}).
(b) Trajectories of a particle pair whose initial orientation is $(\theta_1,\theta_2)=(\pi/2,\pi/16)$ for $\mu=3$ (SP2, blue inverted triangle in Fig.~\ref{fig:phase}).
(c) Trajectories of a particle pair whose initial orientation is $(\theta_1,\theta_2)=(3\pi/4,\pi/2)$ for $\mu=10$.
}
\label{fig:mu3}
\end{figure*}

\section{Results for numerical simulations}
\label{sec:result}

We investigate the dynamics of two active particles by solving the nonlinear equation~(\ref{eq:velocity}) for the absolute positions of the both dipoles and Eqs.~(\ref{eq:r})--(\ref{eq:phi}) for $r$, $\theta_1$, and $\theta_2$.
First, we analyze the active particle trajectories for the three cases ($\eta_{\rm o}=0$, $\eta_{\rm o}\neq0$, and $\eta_{\rm o}\to\infty$), and classify their behaviors into several characteristic states.
Then, we perform numerical simulations with different initial relative angles, and plot the phase diagram for the two angles while keeping the initial inter-particle distance $r$ constant.
To understand the mechanism of the particle collective behavior, we plot the phase space for $\theta_1$ and $\theta_2$ using Eq.~(\ref{eq:theta12}), and perform a linear stability analysis to obtain the characteristic time scale for the pair dynamics.

\subsection{Collective behavior of the active particle pair}
\label{sec:collective}

When $\eta_{\rm o}=0$ (or equivalently, $\mu=0$), we examine the typical temporal evolution of the pair inter-particle distance $r$ for various values of the initial angles, $\theta_1$ and $\theta_2$, while keeping the initial distance fixed at $r=4a$.
When $(\theta_1,\theta_2)=(\pi/2,\pi/2)$, Fig.~\ref{fig:mu0}(a) shows that the two particles approach each other and collide at $r=a$.
For other values of $(\theta_1,\theta_2)$, they move away monotonically and nonmonotonically as shown in Fig.~\ref{fig:mu0}(b) [$(\theta_1,\theta_2)=(\pi/4,\pi/4)$] and Fig.~\ref{fig:mu0}(c) [$(\theta_1,\theta_2)=(\pi/2,\pi/4)$].
We classify these three behaviors as ``convergence state'' (CS), ``monotonic divergence state'' (MDS), and ``nonmonotonic divergence state'' (NDS), shown respectively in Figs.~\ref{fig:mu0}(a), (b), and (c).

Finite values of the odd viscosity $(\mu\neq0)$ result in a perpendicular (nonreciprocal) component $\widehat{\mathbf{d}}^\ast_1$ in the induced velocity, as in Eq.~(\ref{eq:velocity}).
Consequently, it leads to the ``spiral state 1'' (SP1) when $(\theta_1,\theta_2)=(3\pi/4,\pi/2)$, as shown in Fig.~\ref{fig:mu3}(a).
For the angles $(\theta_1,\theta_2)=(\pi/2,\pi/16)$ [Fig.~\ref{fig:mu3}(b)], the active particles show orbiting behavior following attractive or repulsive trajectories before the spiral behavior, which we call the ``spiral state 2'' (SP2).
As the odd viscosity increases, spiral patterns become more evident as shown in Fig.~\ref{fig:mu3}(c) for $\mu=10$.

\begin{figure*}[htb]
\centering
\includegraphics[scale=.55]{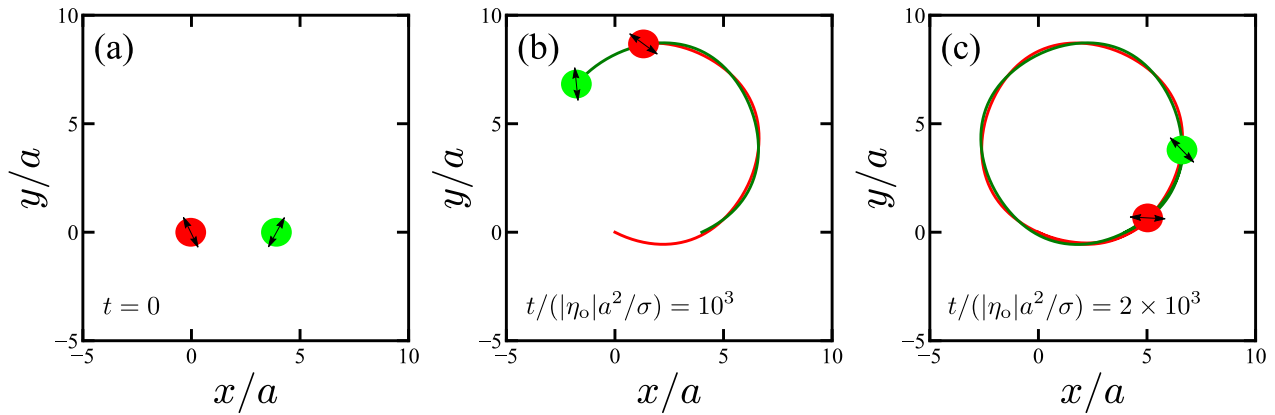}
\caption{
Pair dynamics of active particles in the limit of $\mu\gg1$.
(a) The particles are initially located at $(x, y)=(0,0)$ and $(4a,0)$ with dipolar orientations $\theta_1=0.64\pi$ and $\theta_2=0.35\pi$, respectively.
Snapshots of the particle dynamics at (b) $t/(|\eta_{\rm o}|a^2/\sigma)=10^3$ and (c) $t/(|\eta_{\rm o}|a^2/\sigma)=2\times10^3$.
The pair dynamics corresponds to the circular oscillation state (COS, green circle) in phase space, Fig.~\ref{fig:phase}(c).
}
\label{fig:mu_infinity}
\end{figure*}

\begin{figure}[htb]
\centering
\includegraphics[scale=.5]{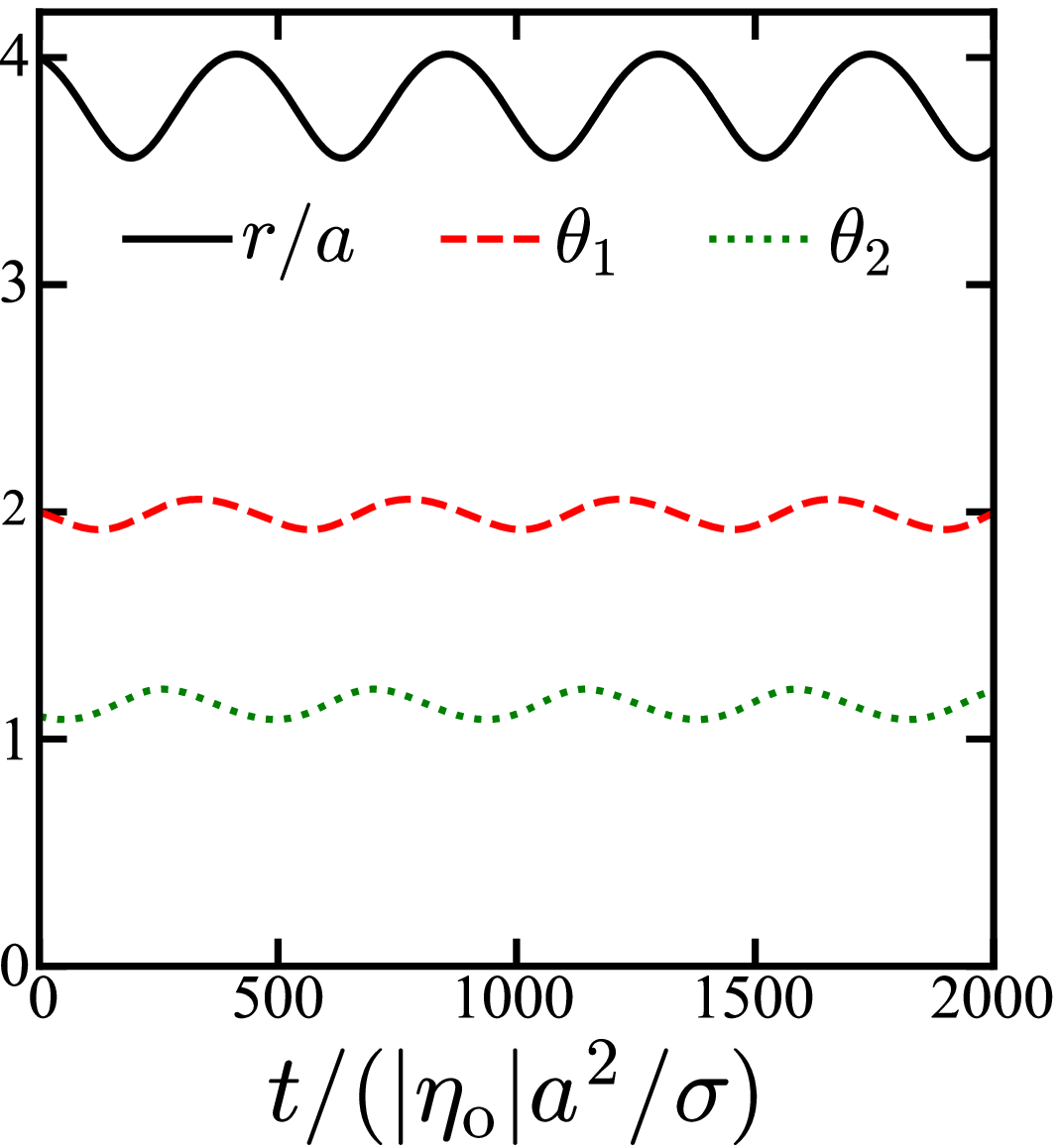}
\caption{
Inter-particle distance $r/a$ of a particle pair (black line) and the two dipolar orientations $\theta_1,\theta_2$ (red dashed and green dotted lines), respectively as a function of the dimensionless time $t/(|\eta_{\rm o}|a^2/\sigma)$.
The initial values are $r/a=4$, $\theta_1=2$, and $\theta_2=1.1$.
}
\label{fig:mu_infinity_rel}
\end{figure}

\begin{figure*}[htb]
\centering
\includegraphics[scale=.55]{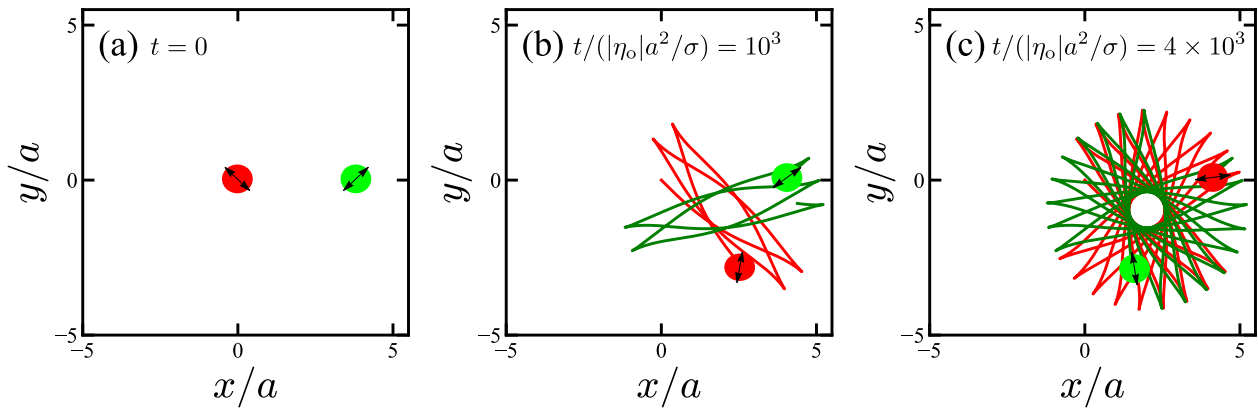}
\caption{
Pair dynamics of active particles in the limit of $\mu\gg1$.
(a) The particles are initially located at $(x, y)=(0,0)$ and $(4a,0)$ with dipolar orientations $\theta_1=3\pi/4$ and $\theta_2=\pi/4$, respectively.
Snapshots of the particle dynamics at (b) $t/(|\eta_{\rm o}|a^2/\sigma)=10^3$ and (c) $t/(|\eta_{\rm o}|a^2/\sigma)=4\times10^3$.
The pair dynamics corresponds to the radial oscillation state (ROS, red star) in the phase space in Fig.~\ref{fig:phase}(c).
}
\label{fig:mu_infinity_star}
\end{figure*}

\begin{figure}[htb]
\centering
\includegraphics[scale=.5]{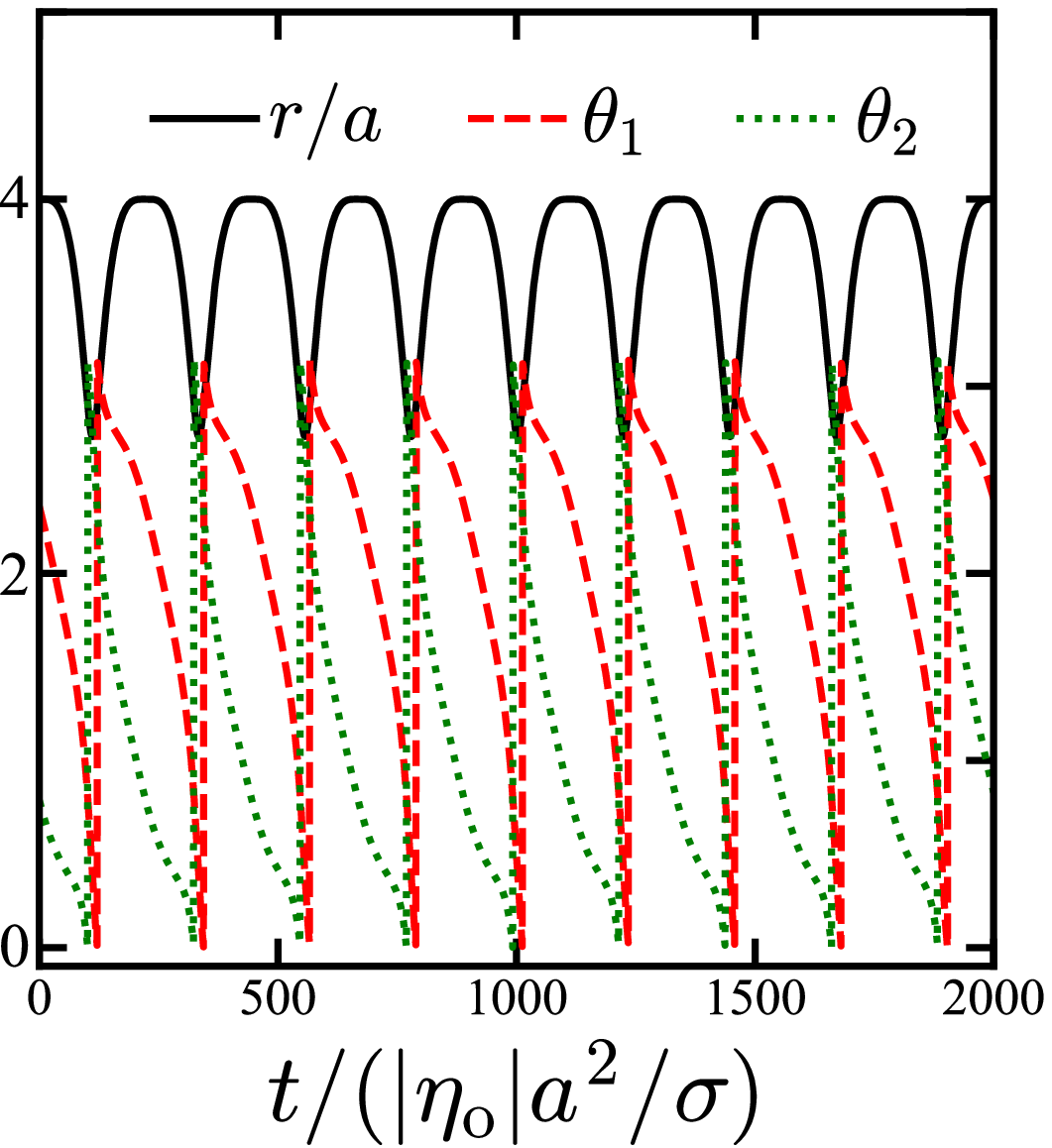}
\caption{
Inter-particle distance $r/a$ of a particle pair (black line), and the two dipolar orientations $\theta_1,\theta_2$ (red dashed and green dotted lines) as a function of the dimensionless time $t/(|\eta_{\rm o}|a^2/\sigma)$.
The initial values are $r/a=4$, $\theta_1=3\pi/4$, and $\theta_2=\pi/4$.
}
\label{fig:mu_infinity_star_rel}
\end{figure}

In the limit of $\mu\gg1$, two types of oscillatory behaviors are seen and they depend on the angles.
Their trajectories and relative dynamics are presented in Figs.~\ref{fig:mu_infinity}, \ref{fig:mu_infinity_rel}, \ref{fig:mu_infinity_star}, and \ref{fig:mu_infinity_star_rel}.
For the dipolar angles $(\theta_1,\theta_2)=(0.64\pi,0.35\pi)$, Fig.~\ref{fig:mu_infinity} shows that one particle follows the other, and they make a circular orbit around a common point located at $(x,y)\approx(2a,4a)$.
Figure~\ref{fig:mu_infinity_rel} shows that the distance and the angles periodically oscillate in time.
This means that the active particles do not remain too close or too far away from each other along the circular trajectory.
For $(\theta_1,\theta_2)=(3\pi/4, \pi/4)$, as shown in Fig.~\ref{fig:mu_infinity_star}, the particles manifest a reciprocal rotation that leads to radial trajectories.
Figure~\ref{fig:mu_infinity_star_rel} exhibits an oscillation period in $r$, $\theta_1$, and $\theta_2$, which is smaller than the period for the circular oscillation shown in Fig.~\ref{fig:mu_infinity_rel}.
We coin the oscillatory behaviors in Figs.~\ref{fig:mu_infinity} and \ref{fig:mu_infinity_star} as the ``circular oscillation state'' (COS) and ``radial oscillation state'' (ROS), respectively.

\subsection{State diagram in the $(\theta_1,\theta_2)$ plane}

\begin{figure*}[htb]
\centering
\includegraphics[scale=.42]{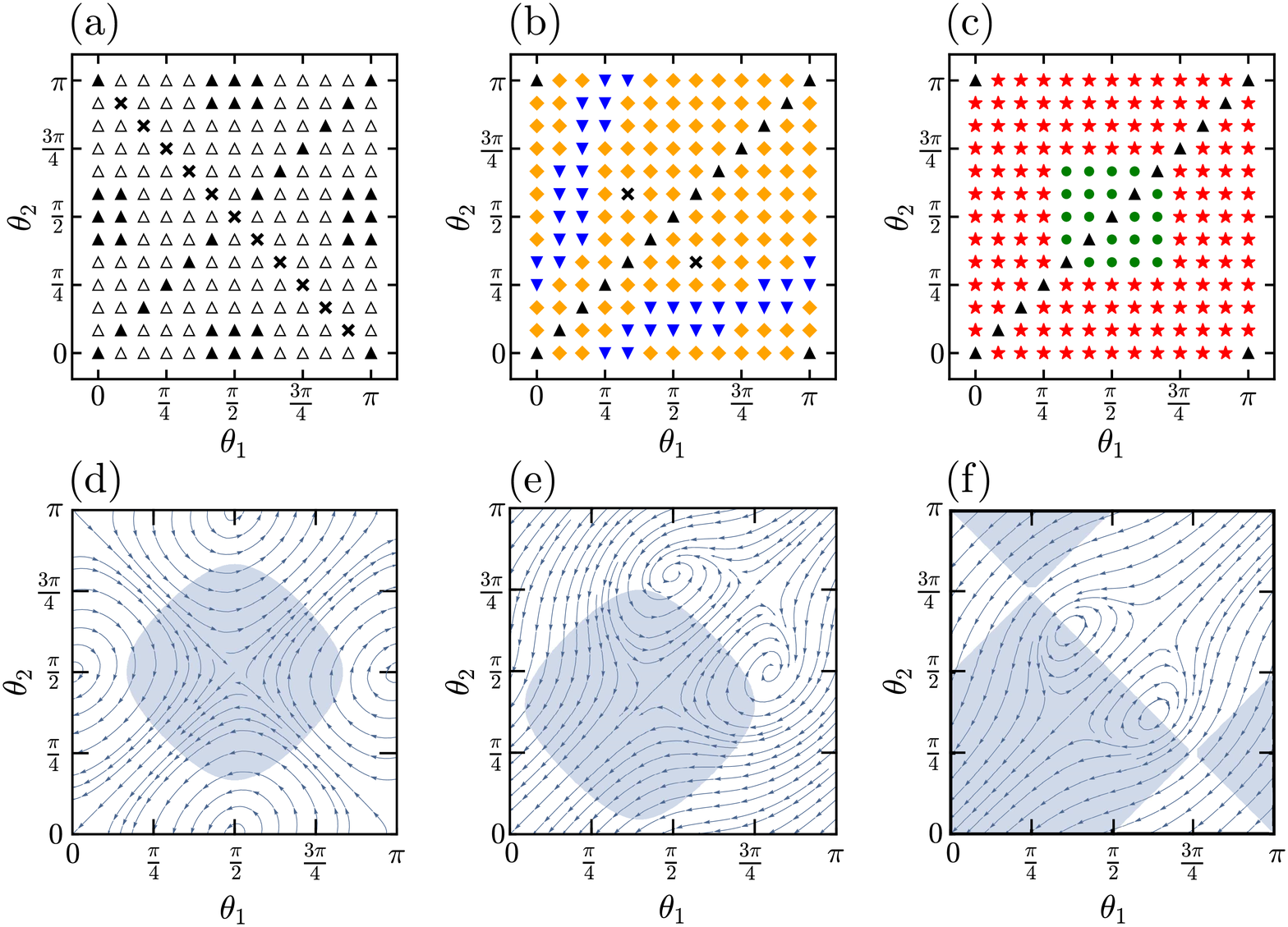}
\caption{
State diagram in the $(\theta_1,\theta_2)$ plane for (a) $\mu=0$, (b) $\mu=3$, and (c) $\mu\gg1$.
Different symbols represent different pair dynamics: monotonic divergence state (MDS, closed triangle), nonmonotonic divergence state (NDS, open triangle), convergence state (CS, cross), spiral state 1 (SP1, orange diamond), spiral state 2 (SP2, blue inverted triangle), circular oscillation state (COS, green circle), and radial oscillation state (ROS, red star).
The phase behavior of $(\theta_1,\theta_2)$ for a fixed value of $r$ for (d) $\mu=0$ [see Eqs.~(\ref{eq:theta1s}) and (\ref{eq:theta2s})], (e) $\mu=3$ [see Eq.~(\ref{eq:theta12})], and (f) $\mu\gg1$ [see Eqs.~(\ref{eq:theta1inf}) and (\ref{eq:theta2inf})].
Shaded regions indicate that the particle pair approaches each other, i.e., when $\dot{r}<0$.
}
\label{fig:phase}
\end{figure*}

To examine the dependence of the active particle dynamics on the polar angles, we numerically integrate Eqs.~(\ref{eq:velocity}) and (\ref{eq:theta12}) with $13$ different initial values of $\theta_1$ for each $\theta_2$.
Figures~\ref{fig:phase}(a), (b), and (c) show the state diagrams in the $(\theta_1,\theta_2)$ plane for $\mu=0$, $3$, and $\mu\gg1$, respectively.
State diagrams are constructed by varying initial angles and categorizing corresponding states of the active particle pair, as shown in Sec.~\ref{sec:collective}.
When $\mu=0$, Fig.~\ref{fig:phase}(a) shows that the particle pair exhibits a ``monotonic divergence state'' (MDS, closed triangle) along the diagonal ($\theta_1=\theta_2$), except for $(\theta_1,\theta_2)=(\pi/2,\pi/2)$.
However, the particle pair along the other diagonal ($\theta_1=-\theta_2$) exhibits a ``convergence state'' (CS, cross).
Note that around the points of $(\theta_1,\theta_2)=(\pi/2,0)$, $(0,\pi/2)$, $(\pi/2,\pi)$, and $(\pi,\pi/2)$, the particles show a MDS behavior (closed triangle), whereas the ``nonmonotonic divergence state'' (NDS, open triangle) is observed for most of the other points in the $(\theta_1,\theta_2)$ plane.

When the odd viscosity is finite, as shown in Fig.~\ref{fig:phase}(b), the ``spiral state 1'' (SP1, orange diamond) and ``spiral state 2'' (SP2, blue inverted triangle) emerge, and the active particle pair shows the two types of spiral behavior, as in Figs.~\ref{fig:mu3}(a) and (b).
In the limit of $\mu\gg1$, as shown in Fig.~\ref{fig:phase}(c), the spiral states disappear altogether and the characteristic circular and radial oscillation states appear in turn.
One can see that around the symmetrical point, $(\theta_1,\theta_2)=(\pi/2,\pi/2)$, the ``circular oscillation state'' (COS, green circle) dominates, while the particles show the ``radial oscillation state'' (ROS, red star) in other points of the $(\theta_1,\theta_2)$ plane.

\subsection{Phase space in dynamical systems}

By connecting the state diagrams with the dynamics of the particle angles, we can analyze more quantitatively the behavior of the active particle pair in the $(\theta_1,\theta_2)$ plane.
The phase space of $\theta_1$ and $\theta_2$ are plotted in Figs.~\ref{fig:phase}(d), (e), and (f) for $\mu=0$, $3$, and $\mu\gg1$, respectively.
The shaded regions in those figure parts indicate that the active particles approach each other ($\dot{r}<0$).
The phase plane for $\mu=0$ is shown in Fig.~\ref{fig:phase}(d), where the vector field ($\dot{\theta}_1,\dot{\theta}_2$) is calculated from Eqs.~(\ref{eq:theta1s}) and (\ref{eq:theta2s}) in Appendix~\ref{app:relative}, while keeping $r$ const.
One can see that there are closed orbits whose centers are located at $(\theta_1,\theta_2)=(\pi/2,0)$, $(0,\pi/2)$, $(\pi/2,\pi)$, and $(\pi,\pi/2)$~\cite{strogatz}.
The point $(\pi/2,\pi/2)$ is unstable along the diagonal line ($\theta_1=\theta_2$), leading to trajectories that move away from that point.
On the other hand, the trajectory starting on the opposite diagonal line ($\theta_1=-\theta_2$) moves towards the $(\pi/2,\pi/2)$ point~\cite{strogatz}.
This point is called a \textit{hyperbolic saddle-point}~\cite{strogatz,manikantan2020}.
Any perturbation of these diagonal points kicks a system into a closed orbit around one of four centers at $(\pi/2,0)$, $(0,\pi/2)$, $(\pi/2,\pi)$, or $(\pi,\pi/2)$ (fixed points)~\cite{strogatz,manikantan2020}.

Using linear stability analysis, we can calculate the time scale of the nonlinear oscillations.
We assume that the angles $\theta_1$ and $\theta_2$ evolve exponentially as $e^{\alpha t}$ when the active particles at a separation $d$ are perturbed from one of the fixed points.
The obtained eigenvalues $\alpha=\pm \sqrt{6}\sigma i/(4\pi\eta_{\rm s}d^2)$ are purely imaginary, leading to an oscillation period $T_0=2\pi/|\alpha|=8\pi^2\eta_{\rm s}d^2/(\sqrt{6}\sigma)$.
Note that most of the initial conditions $\theta_1$, $\theta_2$ in Fig.~\ref{fig:phase}(a) show the divergence state (MDS or NDS) for which $d$ is not constant in time.
Hence, the particles do not show closed orbits characterized by $T_0$.

When $\mu$ is finite, the eigenvalue $\alpha$ has a negative real part, ${\rm Re}(\alpha)<0$, leading to exponentially decaying behavior.
As seen in Fig.~\ref{fig:phase}(e), the fixed points at $(\pi/2,\pi)$ and $(\pi,\pi/2)$ in Fig.~\ref{fig:phase}(d) change to stable spirals that are located approximately at $(\pi/2,3\pi/4)$ and $(3\pi/4,\pi/2)$.
Note that the hyperbolic saddle point remains at ($\pi/2,\pi/2$).
Due to the emerging spirals, most initial values converge to one of the fixed points, leading to constant angles $\theta_1,\theta_2$ in the long-time limit.
Since the spiral centers are not inside the shaded region, particles show the divergence spiral state rather than the convergence spiral state.
Apart from the spirals, some of the initial values do not directly converge to the spiral centers located approximately at $(3\pi/4,\pi/2)$ and $(\pi/2,3\pi/4)$.
Thus, the particles show reciprocating trajectories at an early stage, while they converge to the spirals in the long-time limit with trajectories as in Fig.~\ref{fig:mu3}(b).

When the magnitude of the odd viscosity is large as compared with the shear viscosity $(\mu\to\infty)$, ${\rm Re}(\alpha)\to0$, and the fixed points with closed orbits again emerge at $(\theta_1,\theta_2)\approx(0.37\pi,0.64\pi)$ and $(0.64\pi,0.37\pi)$, as can be seen in Fig.~\ref{fig:phase}(f).
Active particles show the ``monotonic divergence state'' (MDS) along the diagonal line $\theta_1=\theta_2$, and any perturbation from this line kicks the particles away into either a ``circular oscillation state'' (COS, green circle) or ``radial oscillation state'' (ROS, red star).
The centers are characterized by the period $T_\infty=4\sqrt{3}\pi^2|\eta_{\rm o}|d^2/(\sqrt{5}\sigma)\sim|\eta_{\rm o}|$ and the oscillations are governed by the magnitude of odd viscosity.
This oscillation period is now determined by the ratio between the odd-viscous resistance $|\eta_{\rm o}|d$ and the characteristic force due to hydrodynamic interactions $\sigma/d$.

\section{Discussion}
\label{sec:discussion}

When the odd viscosity is zero, the active particle pair mostly shows ``monotonic divergence state'' [MDS, Fig.~\ref{fig:mu0}(b)] and ``nonmonotonic divergence state'' [NDS, Fig.~\ref{fig:mu0}(c)].
Generally speaking, one can see that the azimuthal component $\widehat{\mathbf{d}}$ in Eq.~(\ref{eq:velocity}) due to the induced flow causes the particle pair to unbind and run away from each other.
This leads to the above diverging states.
When the odd viscosity is finite, the particle pair displays spiral states (Fig.~\ref{fig:mu3}) that are due to both the azimuthal $\widehat{\mathbf{d}}_1$ and nonreciprocal components $\widehat{\mathbf{d}}^\ast_1$.
However, when the odd viscosity dominates, the $\widehat{\mathbf{d}}$-term vanishes and the nonreciprocal term $\widehat{\mathbf{d}}^\ast$ causes the pair to exhibit an oscillatory behavior (Figs.~\ref{fig:mu_infinity} and \ref{fig:mu_infinity_star}).
These results demonstrate that the nonreciprocal interaction due to odd viscosity gives rise not only to chiral spiral behavior, but also to oscillatory pair dynamics at sufficiently large values of the odd viscosity.

To see the effect of odd viscosity on the dynamics of a pair of active particles, we compare our results with those for a passive incompressible fluid without odd viscosity~\cite{manikantan2020}.
At length scales smaller than the hydrodynamic screening length, Manikantan showed~\cite{manikantan2020} that pairs of particles display oscillatory dynamics in a 1D coordinate system where one particle follows the other one.
In contrast, our results revealed that the particles show an oscillatory behavior in 2D (as in Figs.~\ref{fig:mu_infinity} and \ref{fig:mu_infinity_star}) due to the nonreciprocal interaction term $\epsilon_{ij}$ in Eq.~(\ref{eq:G}).
On the other hand, at length scales larger than the hydrodynamic screening length, the odd-viscosity term becomes exponentially smaller than the regular viscosity terms, as seen in Eq.~(\ref{eq:Glarge}).
This leads to vanishing chiral and oscillatory dynamics.
Furthermore, clustering behavior due to momentum leakage can emerge, as was observed before~\cite{manikantan2020}.

The inter-particle distance $r$ shown in Fig.~\ref{fig:mu_infinity_rel} represents the oscillatory behavior whose period is determined from the eigenvalue, $T_\infty/(|\eta_{\rm o}|a^2/\sigma)\approx490$, where $d=4a$ is used.
Note that the ``radial oscillation state'' (ROS, shown in Fig.~\ref{fig:mu_infinity_star}) is not determined by $T_\infty$ because the oscillatory behavior is not due to the closed orbits around the centers located approximately at $(0.37\pi, 0.64\pi)$ or $(0.64\pi,0.37\pi)$.
This can be understood by the fact that the area with green circles in Fig.~\ref{fig:phase}(c) almost overlaps with the area around these centers in Fig.~\ref{fig:phase}(f).
Note that such an oscillatory behavior is not observed in passive systems, where the oscillations are governed solely by closed orbits around fixed points in dynamical systems~\cite{manikantan2020}.

It is useful to give some numerical estimates of the physical quantities in our model.
Considering enzymatic molecules, we use typical molecular size $a\approx\SI[parse-numbers=false]{10^{-9}}{\metre}$ 
and estimate the exerted force $f\approx\SI[parse-numbers=false]{10^{-11}}{\newton}$ to obtain 
$\sigma\approx\SI[parse-numbers=false]{10^{-20}}{\newton\metre}$.
Experimental and theoretical findings showed that fluids with spinning particles exhibit an odd viscosity that 
is slightly smaller than the shear viscosity, i.e., $\left|\mu\right|\lesssim1$~\cite{soni2019,hargus2020}.
In such an odd-viscous fluid, one might observe spiral trajectories (see Fig.~\ref{fig:mu3}).

For living systems, however, odd viscosity has not yet been measured, and the possible $\mu$ magnitude can vary significantly depending on the degree of activity.
In the limit of $\mu\gg1$, 
the active particles show circular trajectories (see Fig.~\ref{fig:mu_infinity}) with the period 
$T_\infty\approx3\times\SI[parse-numbers=false]{10^{-5}}{\second}$, where we have 
assumed $d=a$ and $\eta_{\rm o}\approx\SI[parse-numbers=false]{10^{-8}}{\pascal\second\metre}$~\cite{soni2019}.
Since this time scale is comparable to the period of biomolecular chemical reactions catalyzed by fast enzymes such as catalase or urease~\cite{hosaka2020_2}, odd viscosity would affect the chemical reactions through the enzymatic collective behavior.

Experimentally, circular pair dynamics predicted in the limit of $\mu\gg1$ can be investigated by observing 
the response of the surrounding passive inclusions~\cite{svetlizky2021}.
Using optical tweezers, Svetlizky \textit{et al.}\ examined the response of a 2D colloidal suspension to a localized circular motion.
Their result suggests that the circular motion of a pair particle in the presence of odd viscosity can be evaluated by the symmetry breaking in the correlation function of the surrounding passive particles~\cite{svetlizky2021}.

\section{Conclusion and outlook}
\label{sec:conc}

In this paper, we have presented a theory of the dynamics of a pair of active force dipoles in a 2D active chiral fluid characterized by an odd viscosity $\eta_{\rm o}$.
The 2D active chiral fluid is described by a mobility tensor in Eq.~(\ref{eq:G}) with an asymmetric (nonreciprocal) part that is a direct consequence of a finite odd viscosity $(\eta_{\rm o}\neq0)$.
Without the odd viscosity, the particle pair shows both convergence and divergence states, as shown in Fig.~\ref{fig:mu0}.
However, with finite values of the odd viscosity, the particles start to exhibit various chiral pair dynamics, such as spiral (Fig.~\ref{fig:mu3}), circular (Fig.~\ref{fig:mu_infinity}), and radial trajectories (Fig.~\ref{fig:mu_infinity_star}), and this chiral dynamics is due to the antisymmetric (nonreciprocal) component of the mobility tensor.
The circular trajectory can be understood as arising from closed orbits at sufficiently large odd viscosity.
Furthermore, our results demonstrate that the nonreciprocal interaction due to the odd viscosity leads to a wealth of pair dynamics of active particles.

In this work, we have considered a transverse friction coefficient $\zeta_\perp$ between a 2D fluid in contact with an underlying 3D fluid.
In other studies, such transverse momentum leakage from the 2D layer to the 3D fluid beneath it was attributed to an anisotropic substrate~\cite{soni2019} or a Coriolis force in geophysical systems~\cite{tauber2019,tauber2020}.
Here, we show that transverse friction can originate from the odd viscosity of the 3D fluid.
When such an odd viscosity is present in the underlying 3D fluid, an additional term $\eta_{\rm o}^{\rm 3D}\partial_z \nabla^{\rm 3D}\times\mathbf{u}$ will appear in the 3D Stokes equation~\cite{markovich2021,khain2022}, where $\nabla^{\rm 3D}$ is the 3D gradient operator and $\mathbf{u}$ is the 3D fluid velocity.
By employing the lubrication approximation for the 3D fluid, we can obtain the force arising from the 3D odd viscosity on the 2D fluid as $\eta_{\rm o}^{\rm 3D}\mathbf{v}^\ast/h$.
This means that the 3D odd viscosity gives rise to a transverse friction coefficient, $\zeta_\perp=\eta_{\rm o}^{\rm 3D}/h$.
A more detailed discussion of such transverse flows due to odd viscosity will be given elsewhere.

The 2D odd viscosity can vary in the 2D plane, although such an effect was not considered in the present work.
In actual biomembranes, active proteins are often inhomogeneously distributed in the membrane and form active protein-rich domains that are called \textit{lipid rafts}.
The odd viscosity can then be different inside and outside the raft domain~\cite{hosaka2021_2}, and the spatial variation of the odd viscosity needs to be considered.
Some micro-organisms that are known to exhibit directional motions along viscosity gradients are called viscotaxis~\cite{liebchen2018}.
Hence, it will be of value to explore how the nonreciprocal flow field due to odd viscosity gradients couples with the viscotaxis convection.
This interesting question is left for future investigations.

\begin{acknowledgments}

We thank A.\ Kakugo and K.\ Yasuda for fruitful discussions and helpful suggestions.
D.A.\ acknowledges support from the Israel Science Foundation (ISF) under Grant No.\ 213/19 and support by the National Natural Science Foundation of China (NSFC) - ISF joint program under Grant No.\ 3396/19.
S.K.\ acknowledges the support by the National Natural Science Foundation of China  (NSFC)  (Nos.\ 12274098 and 
12250710127) and the startup grant of Wenzhou Institute, University of Chinese Academy of Sciences 
(No.\ WIUCASQD2021041).
\end{acknowledgments}

\section*{Author contribution statement}

The three authors did the research and wrote the article together.

\section*{Data availability}

The data that support the findings of this study are available from the corresponding author upon reasonable request.

\begin{widetext}
\appendix
\section{Derivation of the mobility tensor $G_{ij}[\mathbf{k}]$ in Eq.~(\ref{eq:gmobilityk})}
\label{app:GFT}

We derive the mobility tensor in Fourier space $\mathbf{G}[\mathbf{k}]$ as given by  Eq.~(\ref{eq:gmobilityk}), following a similar derivation in Ref.~\cite{hosaka2021}.
The 2D Fourier transform of $\mathbf{v}(\mathbf{r})$ is defined by
\begin{align}
\mathbf{v}[\mathbf{k}]&=\int {\rm d}^{2} r \, \mathbf{v}(\mathbf{r}) e^{-i \mathbf{k} \cdot \mathbf{r}},
\end{align}
with $\mathbf{k}=(k_x,k_y)$, and the inverse Fourier transform is
\begin{align}
\mathbf{v}(\mathbf{r})&=\int \frac{{\rm d}^{2} k}{(2 \pi)^{2}} \, \mathbf{v}[\mathbf{k}] e^{i \mathbf{k} \cdot \mathbf{r}}.
\end{align}
Similarly, the 2D Fourier transform of the pressure $p(\mathbf{r})$ and force density $\mathbf{F}(\mathbf{r})$ is $p[\mathbf{k}]$ and $\mathbf{F}[\mathbf{k}]$, respectively.
In Fourier space, Eq.~(\ref{eq:stokes}) becomes
\begin{align}
 -\eta_{\rm s}k^2\mathbf{v}[\mathbf{k}]
-\eta_{\rm d} k^2 \widehat{\mathbf{k}} \widehat{\mathbf{k}}\cdot\mathbf{v}[\mathbf{k}]
-\eta_{\rm o} k^2 (\widehat{\mathbf{k}}\overline{\mathbf{k}}\cdot\mathbf{v}[\mathbf{k}]-\overline{\mathbf{k}}\widehat{\mathbf{k}}\cdot\mathbf{v}[\mathbf{k}]) 
 -\frac{ih}{2} k p[\mathbf{k}] \widehat{\mathbf{k}}
-\left(\zeta_\|\mathbf{v}[\mathbf{k}]
+\zeta_\perp\boldsymbol{\epsilon}\cdot\mathbf{v}[\mathbf{k}]
\right)
+ \mathbf{F}[\mathbf{k}] = 0,
\label{eq:monolayerft}
\end{align}
or equivalently
\begin{align}
 -\eta_{\rm s}k^2\mathbf{v}[\mathbf{k}]
-\eta_{\rm d} k^2 v_\|[\mathbf{k}]\widehat{\mathbf{k}} 
-\eta_{\rm o} k^2 (v_\perp[\mathbf{k}]\widehat{\mathbf{k}}-v_\|[\mathbf{k}]\overline{\mathbf{k}}) 
-\frac{ih}{2} k p[\mathbf{k}] \widehat{\mathbf{k}}
-\left(\zeta_\|\mathbf{v}[\mathbf{k}]
+\zeta_\perp\boldsymbol{\epsilon}\cdot\mathbf{v}[\mathbf{k}]
\right)
+ \mathbf{F}[\mathbf{k}] = 0, 
\end{align}
where the two velocity components are $v_\|[\mathbf{k}]=\widehat{\mathbf{k}}\cdot\mathbf{v}[\mathbf{k}]$ and $v_\perp[\mathbf{k}]=\overline{\mathbf{k}}\cdot\mathbf{v}[\mathbf{k}]$.
The compressibility condition, Eq.~(\ref{eq:compress}), becomes
\begin{align}
ik\widehat{\mathbf{k}}\cdot\mathbf{v}[\mathbf{k}] 
=ikv_\|[\mathbf{k}]
= -\frac{h^2}{6\eta}k^2p[\mathbf{k}].
\label{eq:compressft}
\end{align}

Substituting Eq.~(\ref{eq:compressft}) into Eq.~(\ref{eq:monolayerft}) to eliminate $p[\mathbf{k}]$,
we obtain
\begin{align}
-\eta_{\rm s}k^2\mathbf{v}[\mathbf{k}]
-\eta_{\rm d} k^2 v_\|[\mathbf{k}]\widehat{\mathbf{k}}
-\eta_{\rm o} k^2 (v_\perp[\mathbf{k}]\widehat{\mathbf{k}}-v_\|[\mathbf{k}]\overline{\mathbf{k}})
-\frac{3\eta}{h} v_\|[\mathbf{k}] \widehat{\mathbf{k}}
-\left(\zeta_\|\mathbf{v}[\mathbf{k}]
+\zeta_\perp\boldsymbol{\epsilon}\cdot\mathbf{v}[\mathbf{k}]
\right)
+ \mathbf{F}[\mathbf{k}]= 0.
\end{align}
Hence, the force density in Fourier space, $\mathbf{F}[\mathbf{k}]$, is written as
\begin{align}
\displaystyle
	\begin{pmatrix}
		F_\|[\mathbf{k}] \\
		F_\perp[\mathbf{k}]
	\end{pmatrix}
	=
	\begin{pmatrix}
	(\eta_{\rm s}+\eta_{\rm d})k^2 + 3\eta/h +\zeta_\|
	& \eta_{\rm o}k^2 + \zeta_\perp
	\\
	-(\eta_{\rm o}k^2 + \zeta_\perp)&
	 \eta_{\rm s}k^2 + \zeta_\|
	\end{pmatrix}
	\begin{pmatrix}
		v_\|[\mathbf{k}] \\
		v_\perp[\mathbf{k}]
	\end{pmatrix}.
\end{align}
Since the mobility tensor $\mathbf{G}[\mathbf{k}]$ in Fourier space satisfies the relation
$\mathbf{v}[\mathbf{k}]=\mathbf{G}[\mathbf{k}] \cdot \mathbf{F}[\mathbf{k}]$,
we obtain $\mathbf{G}[\mathbf{k}]$ as in Eq.~(\ref{eq:gmobilityk}):
\begin{align}
G_{ij}[\mathbf{k}] =
\frac{
\eta_{\rm s}(k^2+\kappa^2)\widehat{k}_i\widehat{k}_j
+(\eta_{\rm s}+\eta_{\rm d})(k^2+\lambda^2)\overline{k}_i\overline{k}_j
-\eta_{\rm o}(k^2+\nu^2)\epsilon_{i j}
}
{\eta_{\rm s}(\eta_{\rm s}+\eta_{\rm d})\left(k^{2}+\kappa^{2}\right)\left(k^{2}+\lambda^{2}\right)+\eta_{\rm o}^2\left(k^{2}+\nu^2\right)^2}.
\end{align}

\section{Derivation of the mobility tensor $G_{i j}(\mathbf{r})$ in Eq.~(\ref{eq:G})}
\label{app:G}

We obtain $\mathbf{G}(\mathbf{r})$ by performing the inverse Fourier transform of $\mathbf{G}[\mathbf{k}]$, Eq.~(\ref{eq:mobilityk}).
By calculating $G_{ii}$, $G_{ij}\widehat{r}_i\widehat{r}_j$, and $G_{ij}\epsilon_{ij}$~\cite{hosaka2021}, we obtain
\begin{align}
2C_1+C_2 &= \frac{5}{\eta_{\rm s}(4+\mu^2)} \int \frac{\dd{^2k}}{(2\pi)^2}  \frac{1}{k^2+\kappa^{2}}
e^{ikr\cos\varphi}
\nonumber\\
&=  \frac{5}{2\pi\eta_{\rm s}(4+\mu^2)} 
\int_0^\infty \dd{k} \frac{kJ_0(kr)}{k^{2}+\kappa^{2}}\nonumber\\
&= \frac{5K_0(\kappa r)}{2\pi\eta_{\rm s}(4+\mu^2)},
\label{eq:2c1c2}
\end{align}
where $\cos\varphi=\widehat{\mathbf{k}}\cdot\widehat{\mathbf{r}}$ in the integrand, $\varphi$ is the angle between the vectors $\mathbf{k}$ and $\mathbf{r}$, $\mu=\eta_{\rm o}/\eta_{\rm s}$, and $J_n(x)$ and $K_n(x)$ are the Bessel function of the first kind and the modified Bessel function of the second kind, respectively~\cite{abramowitzhandbook}.
In addition,
\begin{align}
C_1+C_2 &= \frac{1}{\eta_{\rm s}(4+\mu^2)} \int \frac{\dd{^2k}}{(2\pi)^2}  
\frac{4-3\cos^2\varphi}{k^{2}+\kappa^{2}}
e^{ikr\cos\varphi}
\nonumber\\
&=  \frac{1}{2\pi\eta_{\rm s}(4+\mu^2)} 
\int_0^\infty \dd{k}   \frac{kr J_0(kr)+3J_1(kr)}{r(k^{2}+\kappa^{2})}\nonumber\\
&= \frac{1}{2\pi\eta_{\rm s}(4+\mu^2)} \left[ \frac{3}{(\kappa r)^2} + K_0(\kappa r) -\frac{3}{\kappa r} \right],
\label{eq:c1c2}
\end{align}
and
\begin{align}
C_3 = -\frac{\mu}{\eta_{\rm s}(4+\mu^2)}  \int \frac{\dd{^2k}}{(2\pi)^2}  \frac{1}{k^2+\kappa^{2}}
e^{ikr\cos\varphi}
=  -\frac{\mu K_0(\kappa r)}{2\pi\eta_{\rm s}(4+\mu^2)}.
\end{align}
From Eqs.~(\ref{eq:2c1c2}) and (\ref{eq:c1c2}), we obtain Eq.~(\ref{eq:c123}) for the three coefficients, $C_1$, $C_2$, and $C_3$.

\section{Derivation of Eqs.~(\ref{eq:velocity}) and (\ref{eq:theta12}) and their asymptotic expressions}
\label{app:relative}

We derive the dynamics of the inter-particle distance $r$ in Eq.~(\ref{eq:velocity}) and the particle polar angles $\theta_1$ and $\theta_2$ in Eq.~(\ref{eq:theta12}).
Substituting Eq.~(\ref{eq:Gsmall}) into Eq.~(\ref{eq:dipole}), we obtain the $i$-component of the in-plane velocity $\mathbf{v}_{21}$ of active particle $2$ relative to $1$ as
\begin{align}
v_{21,i}(\mathbf{r}_{12})
& = -\frac{\sigma_1\widehat{d}_{1,k}\widehat{d}_{1,j} }{4\pi\eta_{\rm s}(4+\mu^2)r}
\Bigl[ 
3(\delta_{ik}\widehat{r}_{12,j}+\delta_{kj}\widehat{r}_{12,i})
-6 \widehat{r}_{12,i}\widehat{r}_{12,j}\widehat{r}_{12,k}
-(5\delta_{ij}-2\mu\epsilon_{ij})\widehat{r}_{12,k}
\Bigl],
\end{align}
where $\sigma_1$ is the force dipole magnitude for particle $1$, $\widehat{\mathbf{d}}_1$ is the dipolar direction of particle $1$, $\delta_{ij}$ is the Kronecker delta, and $\epsilon_{ij}$ is the 2D Levi-Civita tensor.
Through Eq.~(\ref{eq:cos}), we obtain $\mathbf{v}_{21}$ in Eq.~(\ref{eq:velocity})
\begin{align}
\mathbf{v}_{21}(\mathbf{r})= \frac{\sigma_1}{4\pi\eta_{\rm s}(4+\mu^2)r}
\left[
3\cos(2\theta_1)\widehat{\mathbf{r}}_{12}+2\cos\theta_1\widehat{\mathbf{d}}_1-2\mu\cos\theta_1\widehat{\mathbf{d}}_1^\ast
\right],
\end{align}
and similarly for $\mathbf{v}_{12}$.

The two vorticity parameters $(\Omega_1,\Omega_2)$ due to the other particle can be written as
\begin{align}
\begin{aligned}
2\Omega_1=(\nabla\times\mathbf{v}_{12})_z &= 
-\frac{\sigma}{2\pi\eta_{\rm s}(4+\mu^2)r^2}
\Bigl[
4\sin(2\theta_2)+\mu\cos(2\theta_2)
\Bigl],\\
2\Omega_2=(\nabla\times\mathbf{v}_{21})_z &= 
-\frac{\sigma}{2\pi\eta_{\rm s}(4+\mu^2)r^2}
\Bigl[
4\sin(2\theta_1)+\mu\cos(2\theta_1)
\Bigl],
\end{aligned}
\end{align}
where $\Omega_1$ and $\Omega_2$ are scalar for 2D rotation.
The angular rotation $\omega$ due to the azimuthal component of the velocity of Eq.~(\ref{eq:velocity}) can be expressed as 
\begin{align}
\omega = \frac{1}{r} 
\left[
\widehat{\mathbf{r}}\times 
\left((\mathbf{v}_{21}-\mathbf{v}_{12})\cdot\widehat{\mathbf{t}}\widehat{\mathbf{t}}\right)
\right]_z,
\label{eq:omega}
\end{align}
where $\widehat{t}_i=-\epsilon_{ij}\widehat{r}_j$ is a unit vector perpendicular to $\widehat{\mathbf{r}}$, and $\widehat{\mathbf{t}}\widehat{\mathbf{t}}$ is a second-rank tensor.
Notice that the angular rotation $\omega$ is the $z$-component of the right-hand-side of Eq.~(\ref{eq:omega}) as the rotation occurs in 2D.
Since the azimuthal component of the velocity is given by
\begin{align}
(\mathbf{v}_{21}-\mathbf{v}_{12})\cdot\widehat{\mathbf{t}} =
\frac{\sigma}{4\pi\eta_{\rm s}(4+\mu^2)r}
\Bigl[\sin(2\theta_1)+\sin(2\theta_2)+2\mu(\cos^2\theta_1+\cos^2\theta_2)
\Bigl],
\end{align}
the rotation of the inter-particle distance $r$ becomes
\begin{align}
\omega = \frac{\sigma}{4\pi\eta_{\rm s}(4+\mu^2)r^2}
\Bigl[\sin(2\theta_1)+\sin(2\theta_2)+2\mu(\cos^2\theta_1+\cos^2\theta_2)
\Bigl].
\end{align}
We note that the angular rotation due to the azimuthal velocity can be expressed as $\dot{\phi}=d\phi/dt=-\omega$.
Here, the opposite sign of $\omega$ is due to the fact that the angles $\theta_1$ and $\theta_2$ decrease when the inter-particle line rotates in counterclockwise (clockwise) direction for $\omega>0$ ($\omega<0$).
Through the relation $\dot{\theta}_1 = \Omega_1+\dot{\phi}$ and $\dot{\theta}_2 = \Omega_2+\dot{\phi}$, we obtain Eq.~(\ref{eq:theta12}):
\begin{align}
\begin{aligned}
\dot{\theta}_1 &=-\frac{\sigma}{4\pi\eta_{\rm s}(4+\mu^2)r^2}[4\sin(2\theta_2)+\mu\cos(2\theta_2)]+\dot{\phi},
\\
\dot{\theta}_2 &=-\frac{\sigma}{4\pi\eta_{\rm s}(4+\mu^2)r^2}[4\sin(2\theta_1)+\mu\cos(2\theta_1)]+\dot{\phi}.
\end{aligned}
\end{align}

In the limit of $\mu\ll1$, the relative dynamics of Eqs.~(\ref{eq:r}) and (\ref{eq:theta12}) reduces to
\begin{align}
\dot{r} &\approx \frac{\sigma}{8\pi\eta_{\rm s}r} 
\Bigl[1+ 2\cos(2\theta_1)+2\cos(2\theta_2)
-\frac{\mu}{2}
\left[\sin(2\theta_1)+\sin(2\theta_2)\right]
\Bigl],
\label{eq:rs}
\\
\dot{\theta}_1 &\approx-\frac{\sigma}{16\pi\eta_{\rm s}r^2}
\Bigl[\sin(2\theta_1)+5\sin(2\theta_2)
+\frac{\mu}{4}\left[2+\cos(2\theta_1)+2\cos(2\theta_2)
\right]
\Bigl]
,\label{eq:theta1s}\\
\dot{\theta}_2 &\approx-\frac{\sigma}{16\pi\eta_{\rm s} r^2}
\Bigl[
5\sin(2\theta_1)+\sin(2\theta_2)
+\frac{\mu}{4}
\left[2+2\cos(2\theta_1)+\cos(2\theta_2)
\right]
\Bigl].
\label{eq:theta2s}
\end{align}
On the other hand, in the opposite limit of $\mu\gg1$, Eqs.~(\ref{eq:r}) and (\ref{eq:theta12}) can be written as 
\begin{align}
\dot{r} &\approx -\frac{\sigma}{4\pi\eta_{\rm o}r} 
\Bigl[\sin(2\theta_1)+\sin(2\theta_2)\Bigl],\\
\dot{\theta}_1 &\approx-\frac{\sigma}{4\pi\eta_{\rm o}r^2}
\Bigl[2+\cos(2\theta_1)+2\cos(2\theta_2)\Bigl],\label{eq:theta1inf}\\
\dot{\theta}_2 &\approx-\frac{\sigma}{4\pi\eta_{\rm o} r^2}
\Bigl[2+2\cos(2\theta_1)+\cos(2\theta_2)\Bigl],\label{eq:theta2inf}
\end{align}
and are governed solely by the odd viscosity $\eta_{\rm o}$.

\end{widetext}


\begin{thebibliography}{52}%
\makeatletter
\providecommand \@ifxundefined [1]{%
 \@ifx{#1\undefined}
}%
\providecommand \@ifnum [1]{%
 \ifnum #1\expandafter \@firstoftwo
 \else \expandafter \@secondoftwo
 \fi
}%
\providecommand \@ifx [1]{%
 \ifx #1\expandafter \@firstoftwo
 \else \expandafter \@secondoftwo
 \fi
}%
\providecommand \natexlab [1]{#1}%
\providecommand \enquote  [1]{``#1''}%
\providecommand \bibnamefont  [1]{#1}%
\providecommand \bibfnamefont [1]{#1}%
\providecommand \citenamefont [1]{#1}%
\providecommand \href@noop [0]{\@secondoftwo}%
\providecommand \href [0]{\begingroup \@sanitize@url \@href}%
\providecommand \@href[1]{\@@startlink{#1}\@@href}%
\providecommand \@@href[1]{\endgroup#1\@@endlink}%
\providecommand \@sanitize@url [0]{\catcode `\\12\catcode `\$12\catcode
  `\&12\catcode `\#12\catcode `\^12\catcode `\_12\catcode `\%12\relax}%
\providecommand \@@startlink[1]{}%
\providecommand \@@endlink[0]{}%
\providecommand \url  [0]{\begingroup\@sanitize@url \@url }%
\providecommand \@url [1]{\endgroup\@href {#1}{\urlprefix }}%
\providecommand \urlprefix  [0]{URL }%
\providecommand \Eprint [0]{\href }%
\providecommand \doibase [0]{https://doi.org/}%
\providecommand \selectlanguage [0]{\@gobble}%
\providecommand \bibinfo  [0]{\@secondoftwo}%
\providecommand \bibfield  [0]{\@secondoftwo}%
\providecommand \translation [1]{[#1]}%
\providecommand \BibitemOpen [0]{}%
\providecommand \bibitemStop [0]{}%
\providecommand \bibitemNoStop [0]{.\EOS\space}%
\providecommand \EOS [0]{\spacefactor3000\relax}%
\providecommand \BibitemShut  [1]{\csname bibitem#1\endcsname}%
\let\auto@bib@innerbib\@empty
\bibitem [{\citenamefont {Alberts}\ \emph {et~al.}(2008)\citenamefont
  {Alberts}, \citenamefont {Johnson}, \citenamefont {Lewis}, \citenamefont
  {Raff}, \citenamefont {Roberts},\ and\ \citenamefont {Walter}}]{albertsbook}%
  \BibitemOpen
  \bibfield  {author} {\bibinfo {author} {\bibfnamefont {B.}~\bibnamefont
  {Alberts}}, \bibinfo {author} {\bibfnamefont {A.}~\bibnamefont {Johnson}},
  \bibinfo {author} {\bibfnamefont {J.}~\bibnamefont {Lewis}}, \bibinfo
  {author} {\bibfnamefont {M.}~\bibnamefont {Raff}}, \bibinfo {author}
  {\bibfnamefont {K.}~\bibnamefont {Roberts}},\ and\ \bibinfo {author}
  {\bibfnamefont {P.}~\bibnamefont {Walter}},\ }\href@noop {} {\emph {\bibinfo
  {title} {Molecular Biology of the Cell}}},\ \bibinfo {edition} {5th}\ ed.\
  (\bibinfo  {publisher} {Garland Science},\ \bibinfo {address} {New York},\
  \bibinfo {year} {2008})\BibitemShut {NoStop}%
\bibitem [{\citenamefont {Gompper}\ \emph {et~al.}(2020)\citenamefont
  {Gompper}, \citenamefont {Winkler}, \citenamefont {Speck}, \citenamefont
  {Solon}, \citenamefont {Nardini}, \citenamefont {Peruani}, \citenamefont
  {L{\"o}wen}, \citenamefont {Golestanian}, \citenamefont {Kaupp},
  \citenamefont {Alvarez} \emph {et~al.}}]{gompper2020}%
  \BibitemOpen
  \bibfield  {author} {\bibinfo {author} {\bibfnamefont {G.}~\bibnamefont
  {Gompper}}, \bibinfo {author} {\bibfnamefont {R.~G.}\ \bibnamefont
  {Winkler}}, \bibinfo {author} {\bibfnamefont {T.}~\bibnamefont {Speck}},
  \bibinfo {author} {\bibfnamefont {A.}~\bibnamefont {Solon}}, \bibinfo
  {author} {\bibfnamefont {C.}~\bibnamefont {Nardini}}, \bibinfo {author}
  {\bibfnamefont {F.}~\bibnamefont {Peruani}}, \bibinfo {author} {\bibfnamefont
  {H.}~\bibnamefont {L{\"o}wen}}, \bibinfo {author} {\bibfnamefont
  {R.}~\bibnamefont {Golestanian}}, \bibinfo {author} {\bibfnamefont {U.~B.}\
  \bibnamefont {Kaupp}}, \bibinfo {author} {\bibfnamefont {L.}~\bibnamefont
  {Alvarez}}, \emph {et~al.},\ }\bibfield  {title} {\bibinfo {title} {The 2020
  motile active matter roadmap},\ }\href@noop {} {\bibfield  {journal}
  {\bibinfo  {journal} {J. Phys.: Condens. Matter}\ }\textbf {\bibinfo {volume}
  {32}},\ \bibinfo {pages} {193001} (\bibinfo {year} {2020})}\BibitemShut
  {NoStop}%
\bibitem [{\citenamefont {Riedel}\ \emph {et~al.}(2015)\citenamefont {Riedel},
  \citenamefont {Gabizon}, \citenamefont {Wilson}, \citenamefont {Hamadani},
  \citenamefont {Tsekouras}, \citenamefont {Marqusee}, \citenamefont
  {Press{\'e}},\ and\ \citenamefont {Bustamante}}]{riedel2015}%
  \BibitemOpen
  \bibfield  {author} {\bibinfo {author} {\bibfnamefont {C.}~\bibnamefont
  {Riedel}}, \bibinfo {author} {\bibfnamefont {R.}~\bibnamefont {Gabizon}},
  \bibinfo {author} {\bibfnamefont {C.~A.}\ \bibnamefont {Wilson}}, \bibinfo
  {author} {\bibfnamefont {K.}~\bibnamefont {Hamadani}}, \bibinfo {author}
  {\bibfnamefont {K.}~\bibnamefont {Tsekouras}}, \bibinfo {author}
  {\bibfnamefont {S.}~\bibnamefont {Marqusee}}, \bibinfo {author}
  {\bibfnamefont {S.}~\bibnamefont {Press{\'e}}},\ and\ \bibinfo {author}
  {\bibfnamefont {C.}~\bibnamefont {Bustamante}},\ }\bibfield  {title}
  {\bibinfo {title} {The heat released during catalytic turnover enhances the
  diffusion of an enzyme},\ }\href@noop {} {\bibfield  {journal} {\bibinfo
  {journal} {Nature}\ }\textbf {\bibinfo {volume} {517}},\ \bibinfo {pages}
  {227} (\bibinfo {year} {2015})}\BibitemShut {NoStop}%
\bibitem [{\citenamefont {Illien}\ \emph
  {et~al.}(2017{\natexlab{a}})\citenamefont {Illien}, \citenamefont {Zhao},
  \citenamefont {Dey}, \citenamefont {Butler}, \citenamefont {Sen},\ and\
  \citenamefont {Golestanian}}]{illien2017}%
  \BibitemOpen
  \bibfield  {author} {\bibinfo {author} {\bibfnamefont {P.}~\bibnamefont
  {Illien}}, \bibinfo {author} {\bibfnamefont {X.}~\bibnamefont {Zhao}},
  \bibinfo {author} {\bibfnamefont {K.~K.}\ \bibnamefont {Dey}}, \bibinfo
  {author} {\bibfnamefont {P.~J.}\ \bibnamefont {Butler}}, \bibinfo {author}
  {\bibfnamefont {A.}~\bibnamefont {Sen}},\ and\ \bibinfo {author}
  {\bibfnamefont {R.}~\bibnamefont {Golestanian}},\ }\bibfield  {title}
  {\bibinfo {title} {Exothermicity is not a necessary condition for enhanced
  diffusion of enzymes},\ }\href@noop {} {\bibfield  {journal} {\bibinfo
  {journal} {Nano Lett.}\ }\textbf {\bibinfo {volume} {17}},\ \bibinfo {pages}
  {4415} (\bibinfo {year} {2017}{\natexlab{a}})}\BibitemShut {NoStop}%
\bibitem [{\citenamefont {Zhao}\ \emph {et~al.}(2017)\citenamefont {Zhao},
  \citenamefont {Dey}, \citenamefont {Jeganathan}, \citenamefont {Butler},
  \citenamefont {C{\'o}rdova-Figueroa},\ and\ \citenamefont {Sen}}]{zhao2017}%
  \BibitemOpen
  \bibfield  {author} {\bibinfo {author} {\bibfnamefont {X.}~\bibnamefont
  {Zhao}}, \bibinfo {author} {\bibfnamefont {K.~K.}\ \bibnamefont {Dey}},
  \bibinfo {author} {\bibfnamefont {S.}~\bibnamefont {Jeganathan}}, \bibinfo
  {author} {\bibfnamefont {P.~J.}\ \bibnamefont {Butler}}, \bibinfo {author}
  {\bibfnamefont {U.~M.}\ \bibnamefont {C{\'o}rdova-Figueroa}},\ and\ \bibinfo
  {author} {\bibfnamefont {A.}~\bibnamefont {Sen}},\ }\bibfield  {title}
  {\bibinfo {title} {Enhanced diffusion of passive tracers in active enzyme
  solutions},\ }\href@noop {} {\bibfield  {journal} {\bibinfo  {journal} {Nano
  Lett.}\ }\textbf {\bibinfo {volume} {17}},\ \bibinfo {pages} {4807} (\bibinfo
  {year} {2017})}\BibitemShut {NoStop}%
\bibitem [{\citenamefont {Sengupta}\ \emph {et~al.}(2013)\citenamefont
  {Sengupta}, \citenamefont {Dey}, \citenamefont {Muddana}, \citenamefont
  {Tabouillot}, \citenamefont {Ibele}, \citenamefont {Butler},\ and\
  \citenamefont {Sen}}]{sengupta2013}%
  \BibitemOpen
  \bibfield  {author} {\bibinfo {author} {\bibfnamefont {S.}~\bibnamefont
  {Sengupta}}, \bibinfo {author} {\bibfnamefont {K.~K.}\ \bibnamefont {Dey}},
  \bibinfo {author} {\bibfnamefont {H.~S.}\ \bibnamefont {Muddana}}, \bibinfo
  {author} {\bibfnamefont {T.}~\bibnamefont {Tabouillot}}, \bibinfo {author}
  {\bibfnamefont {M.~E.}\ \bibnamefont {Ibele}}, \bibinfo {author}
  {\bibfnamefont {P.~J.}\ \bibnamefont {Butler}},\ and\ \bibinfo {author}
  {\bibfnamefont {A.}~\bibnamefont {Sen}},\ }\bibfield  {title} {\bibinfo
  {title} {Enzyme molecules as nanomotors},\ }\href@noop {} {\bibfield
  {journal} {\bibinfo  {journal} {J. Am. Chem. Soc.}\ }\textbf {\bibinfo
  {volume} {135}},\ \bibinfo {pages} {1406} (\bibinfo {year}
  {2013})}\BibitemShut {NoStop}%
\bibitem [{\citenamefont {Sengupta}\ \emph {et~al.}(2014)\citenamefont
  {Sengupta}, \citenamefont {Spiering}, \citenamefont {Dey}, \citenamefont
  {Duan}, \citenamefont {Patra}, \citenamefont {Butler}, \citenamefont
  {Astumian}, \citenamefont {Benkovic},\ and\ \citenamefont
  {Sen}}]{sengupta2014}%
  \BibitemOpen
  \bibfield  {author} {\bibinfo {author} {\bibfnamefont {S.}~\bibnamefont
  {Sengupta}}, \bibinfo {author} {\bibfnamefont {M.~M.}\ \bibnamefont
  {Spiering}}, \bibinfo {author} {\bibfnamefont {K.~K.}\ \bibnamefont {Dey}},
  \bibinfo {author} {\bibfnamefont {W.}~\bibnamefont {Duan}}, \bibinfo {author}
  {\bibfnamefont {D.}~\bibnamefont {Patra}}, \bibinfo {author} {\bibfnamefont
  {P.~J.}\ \bibnamefont {Butler}}, \bibinfo {author} {\bibfnamefont {R.~D.}\
  \bibnamefont {Astumian}}, \bibinfo {author} {\bibfnamefont {S.~J.}\
  \bibnamefont {Benkovic}},\ and\ \bibinfo {author} {\bibfnamefont
  {A.}~\bibnamefont {Sen}},\ }\bibfield  {title} {\bibinfo {title} {{DNA}
  polymerase as a molecular motor and pump},\ }\href@noop {} {\bibfield
  {journal} {\bibinfo  {journal} {ACS Nano}\ }\textbf {\bibinfo {volume} {8}},\
  \bibinfo {pages} {2410} (\bibinfo {year} {2014})}\BibitemShut {NoStop}%
\bibitem [{\citenamefont {Yu}\ \emph {et~al.}(2009)\citenamefont {Yu},
  \citenamefont {Jo}, \citenamefont {Kounovsky}, \citenamefont {Pablo},\ and\
  \citenamefont {Schwartz}}]{yu2009}%
  \BibitemOpen
  \bibfield  {author} {\bibinfo {author} {\bibfnamefont {H.}~\bibnamefont
  {Yu}}, \bibinfo {author} {\bibfnamefont {K.}~\bibnamefont {Jo}}, \bibinfo
  {author} {\bibfnamefont {K.~L.}\ \bibnamefont {Kounovsky}}, \bibinfo {author}
  {\bibfnamefont {J.~J.~d.}\ \bibnamefont {Pablo}},\ and\ \bibinfo {author}
  {\bibfnamefont {D.~C.}\ \bibnamefont {Schwartz}},\ }\bibfield  {title}
  {\bibinfo {title} {Molecular propulsion: chemical sensing and chemotaxis of
  {DNA} driven by {RNA} polymerase},\ }\href@noop {} {\bibfield  {journal}
  {\bibinfo  {journal} {J. Am. Chem. Soc.}\ }\textbf {\bibinfo {volume}
  {131}},\ \bibinfo {pages} {5722} (\bibinfo {year} {2009})}\BibitemShut
  {NoStop}%
\bibitem [{\citenamefont {Dey}\ \emph {et~al.}(2014)\citenamefont {Dey},
  \citenamefont {Das}, \citenamefont {Poyton}, \citenamefont {Sengupta},
  \citenamefont {Butler}, \citenamefont {Cremer},\ and\ \citenamefont
  {Sen}}]{dey2014}%
  \BibitemOpen
  \bibfield  {author} {\bibinfo {author} {\bibfnamefont {K.~K.}\ \bibnamefont
  {Dey}}, \bibinfo {author} {\bibfnamefont {S.}~\bibnamefont {Das}}, \bibinfo
  {author} {\bibfnamefont {M.~F.}\ \bibnamefont {Poyton}}, \bibinfo {author}
  {\bibfnamefont {S.}~\bibnamefont {Sengupta}}, \bibinfo {author}
  {\bibfnamefont {P.~J.}\ \bibnamefont {Butler}}, \bibinfo {author}
  {\bibfnamefont {P.~S.}\ \bibnamefont {Cremer}},\ and\ \bibinfo {author}
  {\bibfnamefont {A.}~\bibnamefont {Sen}},\ }\bibfield  {title} {\bibinfo
  {title} {Chemotactic separation of enzymes},\ }\href@noop {} {\bibfield
  {journal} {\bibinfo  {journal} {ACS Nano}\ }\textbf {\bibinfo {volume} {8}},\
  \bibinfo {pages} {11941} (\bibinfo {year} {2014})}\BibitemShut {NoStop}%
\bibitem [{\citenamefont {Jee}\ \emph {et~al.}(2018)\citenamefont {Jee},
  \citenamefont {Dutta}, \citenamefont {Cho}, \citenamefont {Tlusty},\ and\
  \citenamefont {Granick}}]{jee2018}%
  \BibitemOpen
  \bibfield  {author} {\bibinfo {author} {\bibfnamefont {A.-Y.}\ \bibnamefont
  {Jee}}, \bibinfo {author} {\bibfnamefont {S.}~\bibnamefont {Dutta}}, \bibinfo
  {author} {\bibfnamefont {Y.-K.}\ \bibnamefont {Cho}}, \bibinfo {author}
  {\bibfnamefont {T.}~\bibnamefont {Tlusty}},\ and\ \bibinfo {author}
  {\bibfnamefont {S.}~\bibnamefont {Granick}},\ }\bibfield  {title} {\bibinfo
  {title} {Enzyme leaps fuel antichemotaxis},\ }\href@noop {} {\bibfield
  {journal} {\bibinfo  {journal} {Proc. Natl. Acad. Sci. (USA)}\ }\textbf
  {\bibinfo {volume} {115}},\ \bibinfo {pages} {14} (\bibinfo {year}
  {2018})}\BibitemShut {NoStop}%
\bibitem [{\citenamefont {Dey}\ \emph {et~al.}(2016)\citenamefont {Dey},
  \citenamefont {Pong}, \citenamefont {Breffke}, \citenamefont {Pavlick},
  \citenamefont {Hatzakis}, \citenamefont {Pacheco},\ and\ \citenamefont
  {Sen}}]{dey2016}%
  \BibitemOpen
  \bibfield  {author} {\bibinfo {author} {\bibfnamefont {K.~K.}\ \bibnamefont
  {Dey}}, \bibinfo {author} {\bibfnamefont {F.~Y.}\ \bibnamefont {Pong}},
  \bibinfo {author} {\bibfnamefont {J.}~\bibnamefont {Breffke}}, \bibinfo
  {author} {\bibfnamefont {R.}~\bibnamefont {Pavlick}}, \bibinfo {author}
  {\bibfnamefont {E.}~\bibnamefont {Hatzakis}}, \bibinfo {author}
  {\bibfnamefont {C.}~\bibnamefont {Pacheco}},\ and\ \bibinfo {author}
  {\bibfnamefont {A.}~\bibnamefont {Sen}},\ }\bibfield  {title} {\bibinfo
  {title} {Dynamic coupling at the {{\AA}}ngstr{\"o}m scale},\ }\href@noop {}
  {\bibfield  {journal} {\bibinfo  {journal} {Angew. Chem.}\ }\textbf {\bibinfo
  {volume} {128}},\ \bibinfo {pages} {1125} (\bibinfo {year}
  {2016})}\BibitemShut {NoStop}%
\bibitem [{\citenamefont {Wang}\ \emph {et~al.}(2020)\citenamefont {Wang},
  \citenamefont {Park}, \citenamefont {Dong}, \citenamefont {Kim},
  \citenamefont {Cho}, \citenamefont {Tlusty},\ and\ \citenamefont
  {Granick}}]{wang2020}%
  \BibitemOpen
  \bibfield  {author} {\bibinfo {author} {\bibfnamefont {H.}~\bibnamefont
  {Wang}}, \bibinfo {author} {\bibfnamefont {M.}~\bibnamefont {Park}}, \bibinfo
  {author} {\bibfnamefont {R.}~\bibnamefont {Dong}}, \bibinfo {author}
  {\bibfnamefont {J.}~\bibnamefont {Kim}}, \bibinfo {author} {\bibfnamefont
  {Y.-K.}\ \bibnamefont {Cho}}, \bibinfo {author} {\bibfnamefont
  {T.}~\bibnamefont {Tlusty}},\ and\ \bibinfo {author} {\bibfnamefont
  {S.}~\bibnamefont {Granick}},\ }\bibfield  {title} {\bibinfo {title} {Boosted
  molecular mobility during common chemical reactions},\ }\href@noop {}
  {\bibfield  {journal} {\bibinfo  {journal} {Science}\ }\textbf {\bibinfo
  {volume} {369}},\ \bibinfo {pages} {537} (\bibinfo {year}
  {2020})}\BibitemShut {NoStop}%
\bibitem [{\citenamefont {MacDonald}\ \emph {et~al.}(2019)\citenamefont
  {MacDonald}, \citenamefont {Price}, \citenamefont {Astumian},\ and\
  \citenamefont {Beves}}]{macdonald2019}%
  \BibitemOpen
  \bibfield  {author} {\bibinfo {author} {\bibfnamefont {T.~S.}\ \bibnamefont
  {MacDonald}}, \bibinfo {author} {\bibfnamefont {W.~S.}\ \bibnamefont
  {Price}}, \bibinfo {author} {\bibfnamefont {R.~D.}\ \bibnamefont
  {Astumian}},\ and\ \bibinfo {author} {\bibfnamefont {J.~E.}\ \bibnamefont
  {Beves}},\ }\bibfield  {title} {\bibinfo {title} {Enhanced diffusion of
  molecular catalysts is due to convection},\ }\href@noop {} {\bibfield
  {journal} {\bibinfo  {journal} {Angew. Chem.}\ }\textbf {\bibinfo {volume}
  {131}},\ \bibinfo {pages} {19040} (\bibinfo {year} {2019})}\BibitemShut
  {NoStop}%
\bibitem [{\citenamefont {Rezaei-Ghaleh}\ \emph {et~al.}(2022)\citenamefont
  {Rezaei-Ghaleh}, \citenamefont {Agudo-Canalejo}, \citenamefont {Griesinger},\
  and\ \citenamefont {Golestanian}}]{rezaei2022}%
  \BibitemOpen
  \bibfield  {author} {\bibinfo {author} {\bibfnamefont {N.}~\bibnamefont
  {Rezaei-Ghaleh}}, \bibinfo {author} {\bibfnamefont {J.}~\bibnamefont
  {Agudo-Canalejo}}, \bibinfo {author} {\bibfnamefont {C.}~\bibnamefont
  {Griesinger}},\ and\ \bibinfo {author} {\bibfnamefont {R.}~\bibnamefont
  {Golestanian}},\ }\bibfield  {title} {\bibinfo {title} {Molecular diffusivity
  of click reaction components: The diffusion enhancement question},\
  }\href@noop {} {\bibfield  {journal} {\bibinfo  {journal} {J. Am. Chem.
  Soc.}\ }\textbf {\bibinfo {volume} {144}},\ \bibinfo {pages} {1380} (\bibinfo
  {year} {2022})}\BibitemShut {NoStop}%
\bibitem [{\citenamefont {Togashi}\ and\ \citenamefont
  {Mikhailov}(2007)}]{togashi2007}%
  \BibitemOpen
  \bibfield  {author} {\bibinfo {author} {\bibfnamefont {Y.}~\bibnamefont
  {Togashi}}\ and\ \bibinfo {author} {\bibfnamefont {A.~S.}\ \bibnamefont
  {Mikhailov}},\ }\href@noop {} {\bibfield  {journal} {\bibinfo  {journal}
  {Proc. Natl. Acad. Sci. (USA)}\ }\textbf {\bibinfo {volume} {104}},\ \bibinfo
  {pages} {8697} (\bibinfo {year} {2007})}\BibitemShut {NoStop}%
\bibitem [{\citenamefont {Illien}\ \emph
  {et~al.}(2017{\natexlab{b}})\citenamefont {Illien}, \citenamefont
  {Adeleke-Larodo},\ and\ \citenamefont {Golestanian}}]{illien2017_epl}%
  \BibitemOpen
  \bibfield  {author} {\bibinfo {author} {\bibfnamefont {P.}~\bibnamefont
  {Illien}}, \bibinfo {author} {\bibfnamefont {T.}~\bibnamefont
  {Adeleke-Larodo}},\ and\ \bibinfo {author} {\bibfnamefont {R.}~\bibnamefont
  {Golestanian}},\ }\bibfield  {title} {\bibinfo {title} {Diffusion of an
  enzyme: The role of fluctuation-induced hydrodynamic coupling},\ }\href@noop
  {} {\bibfield  {journal} {\bibinfo  {journal} {EPL}\ }\textbf {\bibinfo
  {volume} {119}},\ \bibinfo {pages} {40002} (\bibinfo {year}
  {2017}{\natexlab{b}})}\BibitemShut {NoStop}%
\bibitem [{\citenamefont {Hosaka}\ \emph
  {et~al.}(2020{\natexlab{a}})\citenamefont {Hosaka}, \citenamefont {Komura},\
  and\ \citenamefont {Andelman}}]{hosaka2020}%
  \BibitemOpen
  \bibfield  {author} {\bibinfo {author} {\bibfnamefont {Y.}~\bibnamefont
  {Hosaka}}, \bibinfo {author} {\bibfnamefont {S.}~\bibnamefont {Komura}},\
  and\ \bibinfo {author} {\bibfnamefont {D.}~\bibnamefont {Andelman}},\
  }\bibfield  {title} {\bibinfo {title} {Shear viscosity of two-state enzyme
  solutions},\ }\href@noop {} {\bibfield  {journal} {\bibinfo  {journal} {Phys.
  Rev. E}\ }\textbf {\bibinfo {volume} {101}},\ \bibinfo {pages} {012610}
  (\bibinfo {year} {2020}{\natexlab{a}})}\BibitemShut {NoStop}%
\bibitem [{\citenamefont {Hosaka}\ \emph
  {et~al.}(2020{\natexlab{b}})\citenamefont {Hosaka}, \citenamefont {Komura},\
  and\ \citenamefont {Mikhailov}}]{hosaka2020_2}%
  \BibitemOpen
  \bibfield  {author} {\bibinfo {author} {\bibfnamefont {Y.}~\bibnamefont
  {Hosaka}}, \bibinfo {author} {\bibfnamefont {S.}~\bibnamefont {Komura}},\
  and\ \bibinfo {author} {\bibfnamefont {A.~S.}\ \bibnamefont {Mikhailov}},\
  }\bibfield  {title} {\bibinfo {title} {Mechanochemical enzymes and protein
  machines as hydrodynamic force dipoles: the active dimer model},\ }\href@noop
  {} {\bibfield  {journal} {\bibinfo  {journal} {Soft Matter}\ }\textbf
  {\bibinfo {volume} {16}},\ \bibinfo {pages} {10734} (\bibinfo {year}
  {2020}{\natexlab{b}})}\BibitemShut {NoStop}%
\bibitem [{\citenamefont {Hosaka}\ \emph {et~al.}(2017)\citenamefont {Hosaka},
  \citenamefont {Yasuda}, \citenamefont {Okamoto},\ and\ \citenamefont
  {Komura}}]{hosaka2017}%
  \BibitemOpen
  \bibfield  {author} {\bibinfo {author} {\bibfnamefont {Y.}~\bibnamefont
  {Hosaka}}, \bibinfo {author} {\bibfnamefont {K.}~\bibnamefont {Yasuda}},
  \bibinfo {author} {\bibfnamefont {R.}~\bibnamefont {Okamoto}},\ and\ \bibinfo
  {author} {\bibfnamefont {S.}~\bibnamefont {Komura}},\ }\bibfield  {title}
  {\bibinfo {title} {Lateral diffusion induced by active proteins in a
  biomembrane},\ }\href@noop {} {\bibfield  {journal} {\bibinfo  {journal}
  {Phys. Rev. E}\ }\textbf {\bibinfo {volume} {95}},\ \bibinfo {pages} {052407}
  (\bibinfo {year} {2017})}\BibitemShut {NoStop}%
\bibitem [{\citenamefont {Mikhailov}\ and\ \citenamefont
  {Kapral}(2015)}]{mikhailov2015}%
  \BibitemOpen
  \bibfield  {author} {\bibinfo {author} {\bibfnamefont {A.~S.}\ \bibnamefont
  {Mikhailov}}\ and\ \bibinfo {author} {\bibfnamefont {R.}~\bibnamefont
  {Kapral}},\ }\bibfield  {title} {\bibinfo {title} {Hydrodynamic collective
  effects of active protein machines in solution and lipid bilayers},\
  }\href@noop {} {\bibfield  {journal} {\bibinfo  {journal} {Proc. Natl. Acad.
  Sci. (USA)}\ }\textbf {\bibinfo {volume} {112}},\ \bibinfo {pages} {E3639}
  (\bibinfo {year} {2015})}\BibitemShut {NoStop}%
\bibitem [{\citenamefont {Manikantan}(2020)}]{manikantan2020}%
  \BibitemOpen
  \bibfield  {author} {\bibinfo {author} {\bibfnamefont {H.}~\bibnamefont
  {Manikantan}},\ }\bibfield  {title} {\bibinfo {title} {Tunable collective
  dynamics of active inclusions in viscous membranes},\ }\href@noop {}
  {\bibfield  {journal} {\bibinfo  {journal} {Phys. Rev. Lett.}\ }\textbf
  {\bibinfo {volume} {125}},\ \bibinfo {pages} {268101} (\bibinfo {year}
  {2020})}\BibitemShut {NoStop}%
\bibitem [{\citenamefont {Bagaria}\ and\ \citenamefont
  {Samanta}(2022)}]{bagaria2022}%
  \BibitemOpen
  \bibfield  {author} {\bibinfo {author} {\bibfnamefont {S.}~\bibnamefont
  {Bagaria}}\ and\ \bibinfo {author} {\bibfnamefont {R.}~\bibnamefont
  {Samanta}},\ }\bibfield  {title} {\bibinfo {title} {Dynamics of force dipoles
  in curved biological membranes},\ }\href@noop {} {\bibfield  {journal}
  {\bibinfo  {journal} {Phys. Rev. Fluids}\ }\textbf {\bibinfo {volume} {7}},\
  \bibinfo {pages} {093101} (\bibinfo {year} {2022})}\BibitemShut {NoStop}%
\bibitem [{\citenamefont {Oppenheimer}\ \emph {et~al.}(2019)\citenamefont
  {Oppenheimer}, \citenamefont {Stein},\ and\ \citenamefont
  {Shelley}}]{oppenheimer2019}%
  \BibitemOpen
  \bibfield  {author} {\bibinfo {author} {\bibfnamefont {N.}~\bibnamefont
  {Oppenheimer}}, \bibinfo {author} {\bibfnamefont {D.~B.}\ \bibnamefont
  {Stein}},\ and\ \bibinfo {author} {\bibfnamefont {M.~J.}\ \bibnamefont
  {Shelley}},\ }\bibfield  {title} {\bibinfo {title} {Rotating membrane
  inclusions crystallize through hydrodynamic and steric interactions},\
  }\href@noop {} {\bibfield  {journal} {\bibinfo  {journal} {Phys. Rev. Lett.}\
  }\textbf {\bibinfo {volume} {123}},\ \bibinfo {pages} {148101} (\bibinfo
  {year} {2019})}\BibitemShut {NoStop}%
\bibitem [{\citenamefont {Manneville}\ \emph {et~al.}(1999)\citenamefont
  {Manneville}, \citenamefont {Bassereau}, \citenamefont {Levy},\ and\
  \citenamefont {Prost}}]{manneville1999}%
  \BibitemOpen
  \bibfield  {author} {\bibinfo {author} {\bibfnamefont {J.-B.}\ \bibnamefont
  {Manneville}}, \bibinfo {author} {\bibfnamefont {P.}~\bibnamefont
  {Bassereau}}, \bibinfo {author} {\bibfnamefont {D.}~\bibnamefont {Levy}},\
  and\ \bibinfo {author} {\bibfnamefont {J.}~\bibnamefont {Prost}},\ }\bibfield
   {title} {\bibinfo {title} {Activity of transmembrane proteins induces
  magnification of shape fluctuations of lipid membranes},\ }\href@noop {}
  {\bibfield  {journal} {\bibinfo  {journal} {Phys. Rev. Lett.}\ }\textbf
  {\bibinfo {volume} {82}},\ \bibinfo {pages} {4356} (\bibinfo {year}
  {1999})}\BibitemShut {NoStop}%
\bibitem [{\citenamefont {Manneville}\ \emph {et~al.}(2001)\citenamefont
  {Manneville}, \citenamefont {Bassereau}, \citenamefont {Ramaswamy},\ and\
  \citenamefont {Prost}}]{manneville2001}%
  \BibitemOpen
  \bibfield  {author} {\bibinfo {author} {\bibfnamefont {J.-B.}\ \bibnamefont
  {Manneville}}, \bibinfo {author} {\bibfnamefont {P.}~\bibnamefont
  {Bassereau}}, \bibinfo {author} {\bibfnamefont {S.}~\bibnamefont
  {Ramaswamy}},\ and\ \bibinfo {author} {\bibfnamefont {J.}~\bibnamefont
  {Prost}},\ }\bibfield  {title} {\bibinfo {title} {Active membrane
  fluctuations studied by micropipet aspiration},\ }\href@noop {} {\bibfield
  {journal} {\bibinfo  {journal} {Phys. Rev. E}\ }\textbf {\bibinfo {volume}
  {64}},\ \bibinfo {pages} {021908} (\bibinfo {year} {2001})}\BibitemShut
  {NoStop}%
\bibitem [{\citenamefont {Banerjee}\ \emph {et~al.}(2017)\citenamefont
  {Banerjee}, \citenamefont {Souslov}, \citenamefont {Abanov},\ and\
  \citenamefont {Vitelli}}]{banerjee2017}%
  \BibitemOpen
  \bibfield  {author} {\bibinfo {author} {\bibfnamefont {D.}~\bibnamefont
  {Banerjee}}, \bibinfo {author} {\bibfnamefont {A.}~\bibnamefont {Souslov}},
  \bibinfo {author} {\bibfnamefont {A.~G.}\ \bibnamefont {Abanov}},\ and\
  \bibinfo {author} {\bibfnamefont {V.}~\bibnamefont {Vitelli}},\ }\bibfield
  {title} {\bibinfo {title} {Odd viscosity in chiral active fluids},\
  }\href@noop {} {\bibfield  {journal} {\bibinfo  {journal} {Nat. Commun.}\
  }\textbf {\bibinfo {volume} {8}},\ \bibinfo {pages} {1573} (\bibinfo {year}
  {2017})}\BibitemShut {NoStop}%
\bibitem [{\citenamefont {Hosaka}\ and\ \citenamefont
  {Komura}(2022)}]{hosaka2022}%
  \BibitemOpen
  \bibfield  {author} {\bibinfo {author} {\bibfnamefont {Y.}~\bibnamefont
  {Hosaka}}\ and\ \bibinfo {author} {\bibfnamefont {S.}~\bibnamefont
  {Komura}},\ }\bibfield  {title} {\bibinfo {title} {Nonequilibrium transport
  induced by biological nanomachines},\ }\href@noop {} {\bibfield  {journal}
  {\bibinfo  {journal} {Biophys. Rev. Lett.}\ }\textbf {\bibinfo {volume}
  {17}},\ \bibinfo {pages} {51} (\bibinfo {year} {2022})}\BibitemShut {NoStop}%
\bibitem [{\citenamefont {Fruchart}\ \emph {et~al.}(2022)\citenamefont
  {Fruchart}, \citenamefont {Scheibner},\ and\ \citenamefont
  {Vitelli}}]{fruchart2022}%
  \BibitemOpen
  \bibfield  {author} {\bibinfo {author} {\bibfnamefont {M.}~\bibnamefont
  {Fruchart}}, \bibinfo {author} {\bibfnamefont {C.}~\bibnamefont
  {Scheibner}},\ and\ \bibinfo {author} {\bibfnamefont {V.}~\bibnamefont
  {Vitelli}},\ }\bibfield  {title} {\bibinfo {title} {Odd viscosity and odd
  elasticity},\ }\href@noop {} {\bibfield  {journal} {\bibinfo  {journal}
  {arXiv preprint arXiv:2207.00071}\ } (\bibinfo {year} {2022})}\BibitemShut
  {NoStop}%
\bibitem [{\citenamefont {Ganeshan}\ and\ \citenamefont
  {Abanov}(2017)}]{ganeshan2017}%
  \BibitemOpen
  \bibfield  {author} {\bibinfo {author} {\bibfnamefont {S.}~\bibnamefont
  {Ganeshan}}\ and\ \bibinfo {author} {\bibfnamefont {A.~G.}\ \bibnamefont
  {Abanov}},\ }\bibfield  {title} {\bibinfo {title} {Odd viscosity in
  two-dimensional incompressible fluids},\ }\href@noop {} {\bibfield  {journal}
  {\bibinfo  {journal} {Phys. Rev. Fluids}\ }\textbf {\bibinfo {volume} {2}},\
  \bibinfo {pages} {094101} (\bibinfo {year} {2017})}\BibitemShut {NoStop}%
\bibitem [{\citenamefont {Souslov}\ \emph {et~al.}(2020)\citenamefont
  {Souslov}, \citenamefont {Gromov},\ and\ \citenamefont
  {Vitelli}}]{souslov2020}%
  \BibitemOpen
  \bibfield  {author} {\bibinfo {author} {\bibfnamefont {A.}~\bibnamefont
  {Souslov}}, \bibinfo {author} {\bibfnamefont {A.}~\bibnamefont {Gromov}},\
  and\ \bibinfo {author} {\bibfnamefont {V.}~\bibnamefont {Vitelli}},\
  }\bibfield  {title} {\bibinfo {title} {Anisotropic odd viscosity via a
  time-modulated drive},\ }\href@noop {} {\bibfield  {journal} {\bibinfo
  {journal} {Phys. Rev. E}\ }\textbf {\bibinfo {volume} {101}},\ \bibinfo
  {pages} {052606} (\bibinfo {year} {2020})}\BibitemShut {NoStop}%
\bibitem [{\citenamefont {Hosaka}\ \emph
  {et~al.}(2021{\natexlab{a}})\citenamefont {Hosaka}, \citenamefont {Komura},\
  and\ \citenamefont {Andelman}}]{hosaka2021}%
  \BibitemOpen
  \bibfield  {author} {\bibinfo {author} {\bibfnamefont {Y.}~\bibnamefont
  {Hosaka}}, \bibinfo {author} {\bibfnamefont {S.}~\bibnamefont {Komura}},\
  and\ \bibinfo {author} {\bibfnamefont {D.}~\bibnamefont {Andelman}},\
  }\bibfield  {title} {\bibinfo {title} {Nonreciprocal response of a
  two-dimensional fluid with odd viscosity},\ }\href@noop {} {\bibfield
  {journal} {\bibinfo  {journal} {Phys. Rev. E}\ }\textbf {\bibinfo {volume}
  {103}},\ \bibinfo {pages} {042610} (\bibinfo {year}
  {2021}{\natexlab{a}})}\BibitemShut {NoStop}%
\bibitem [{\citenamefont {Hosaka}\ \emph
  {et~al.}(2021{\natexlab{b}})\citenamefont {Hosaka}, \citenamefont {Komura},\
  and\ \citenamefont {Andelman}}]{hosaka2021_2}%
  \BibitemOpen
  \bibfield  {author} {\bibinfo {author} {\bibfnamefont {Y.}~\bibnamefont
  {Hosaka}}, \bibinfo {author} {\bibfnamefont {S.}~\bibnamefont {Komura}},\
  and\ \bibinfo {author} {\bibfnamefont {D.}~\bibnamefont {Andelman}},\
  }\bibfield  {title} {\bibinfo {title} {Hydrodynamic lift of a two-dimensional
  liquid domain with odd viscosity},\ }\href@noop {} {\bibfield  {journal}
  {\bibinfo  {journal} {Phys. Rev. E}\ }\textbf {\bibinfo {volume} {104}},\
  \bibinfo {pages} {064613} (\bibinfo {year} {2021}{\natexlab{b}})}\BibitemShut
  {NoStop}%
\bibitem [{\citenamefont {Lier}\ \emph {et~al.}(2022)\citenamefont {Lier},
  \citenamefont {Duclut}, \citenamefont {Bo}, \citenamefont {Armas},
  \citenamefont {J{\"u}licher},\ and\ \citenamefont {Sur{\'o}wka}}]{lier2022}%
  \BibitemOpen
  \bibfield  {author} {\bibinfo {author} {\bibfnamefont {R.}~\bibnamefont
  {Lier}}, \bibinfo {author} {\bibfnamefont {C.}~\bibnamefont {Duclut}},
  \bibinfo {author} {\bibfnamefont {S.}~\bibnamefont {Bo}}, \bibinfo {author}
  {\bibfnamefont {J.}~\bibnamefont {Armas}}, \bibinfo {author} {\bibfnamefont
  {F.}~\bibnamefont {J{\"u}licher}},\ and\ \bibinfo {author} {\bibfnamefont
  {P.}~\bibnamefont {Sur{\'o}wka}},\ }\bibfield  {title} {\bibinfo {title}
  {Lift force in odd compressible fluids},\ }\href@noop {} {\bibfield
  {journal} {\bibinfo  {journal} {arXiv preprint arXiv:2205.12704}\ } (\bibinfo
  {year} {2022})}\BibitemShut {NoStop}%
\bibitem [{\citenamefont {Khain}\ \emph {et~al.}(2022)\citenamefont {Khain},
  \citenamefont {Scheibner}, \citenamefont {Fruchart},\ and\ \citenamefont
  {Vitelli}}]{khain2022}%
  \BibitemOpen
  \bibfield  {author} {\bibinfo {author} {\bibfnamefont {T.}~\bibnamefont
  {Khain}}, \bibinfo {author} {\bibfnamefont {C.}~\bibnamefont {Scheibner}},
  \bibinfo {author} {\bibfnamefont {M.}~\bibnamefont {Fruchart}},\ and\
  \bibinfo {author} {\bibfnamefont {V.}~\bibnamefont {Vitelli}},\ }\bibfield
  {title} {\bibinfo {title} {Stokes flows in three-dimensional fluids with odd
  and parity-violating viscosities},\ }\href@noop {} {\bibfield  {journal}
  {\bibinfo  {journal} {J. Fluid Mech.}\ }\textbf {\bibinfo {volume} {934}},\
  \bibinfo {pages} {A23} (\bibinfo {year} {2022})}\BibitemShut {NoStop}%
\bibitem [{\citenamefont {Markovich}\ and\ \citenamefont
  {Lubensky}(2021)}]{markovich2021}%
  \BibitemOpen
  \bibfield  {author} {\bibinfo {author} {\bibfnamefont {T.}~\bibnamefont
  {Markovich}}\ and\ \bibinfo {author} {\bibfnamefont {T.~C.}\ \bibnamefont
  {Lubensky}},\ }\bibfield  {title} {\bibinfo {title} {Odd viscosity in active
  matter: Microscopic origin and 3{D} effects},\ }\href@noop {} {\bibfield
  {journal} {\bibinfo  {journal} {Phys. Rev. Lett.}\ }\textbf {\bibinfo
  {volume} {127}},\ \bibinfo {pages} {048001} (\bibinfo {year}
  {2021})}\BibitemShut {NoStop}%
\bibitem [{\citenamefont {Barentin}\ \emph {et~al.}(1999)\citenamefont
  {Barentin}, \citenamefont {Ybert}, \citenamefont {di~Meglio},\ and\
  \citenamefont {Joanny}}]{barentin1999}%
  \BibitemOpen
  \bibfield  {author} {\bibinfo {author} {\bibfnamefont {C.}~\bibnamefont
  {Barentin}}, \bibinfo {author} {\bibfnamefont {C.}~\bibnamefont {Ybert}},
  \bibinfo {author} {\bibfnamefont {J.-M.}\ \bibnamefont {di~Meglio}},\ and\
  \bibinfo {author} {\bibfnamefont {J.-F.}\ \bibnamefont {Joanny}},\ }\bibfield
   {title} {\bibinfo {title} {Surface shear viscosity of gibbs and langmuir
  monolayers},\ }\href@noop {} {\bibfield  {journal} {\bibinfo  {journal} {J.
  Fluid Mech.}\ }\textbf {\bibinfo {volume} {397}},\ \bibinfo {pages} {331}
  (\bibinfo {year} {1999})}\BibitemShut {NoStop}%
\bibitem [{\citenamefont {Elfring}\ \emph {et~al.}(2016)\citenamefont
  {Elfring}, \citenamefont {Leal},\ and\ \citenamefont
  {Squires}}]{elfring2016}%
  \BibitemOpen
  \bibfield  {author} {\bibinfo {author} {\bibfnamefont {G.~J.}\ \bibnamefont
  {Elfring}}, \bibinfo {author} {\bibfnamefont {L.~G.}\ \bibnamefont {Leal}},\
  and\ \bibinfo {author} {\bibfnamefont {T.~M.}\ \bibnamefont {Squires}},\
  }\bibfield  {title} {\bibinfo {title} {Surface viscosity and marangoni
  stresses at surfactant laden interfaces},\ }\href@noop {} {\bibfield
  {journal} {\bibinfo  {journal} {J. Fluid Mech.}\ }\textbf {\bibinfo {volume}
  {792}},\ \bibinfo {pages} {712} (\bibinfo {year} {2016})}\BibitemShut
  {NoStop}%
\bibitem [{\citenamefont {Soni}\ \emph {et~al.}(2019)\citenamefont {Soni},
  \citenamefont {Bililign}, \citenamefont {Magkiriadou}, \citenamefont
  {Sacanna}, \citenamefont {Bartolo}, \citenamefont {Shelley},\ and\
  \citenamefont {Irvine}}]{soni2019}%
  \BibitemOpen
  \bibfield  {author} {\bibinfo {author} {\bibfnamefont {V.}~\bibnamefont
  {Soni}}, \bibinfo {author} {\bibfnamefont {E.~S.}\ \bibnamefont {Bililign}},
  \bibinfo {author} {\bibfnamefont {S.}~\bibnamefont {Magkiriadou}}, \bibinfo
  {author} {\bibfnamefont {S.}~\bibnamefont {Sacanna}}, \bibinfo {author}
  {\bibfnamefont {D.}~\bibnamefont {Bartolo}}, \bibinfo {author} {\bibfnamefont
  {M.~J.}\ \bibnamefont {Shelley}},\ and\ \bibinfo {author} {\bibfnamefont
  {W.~T.~M.}\ \bibnamefont {Irvine}},\ }\bibfield  {title} {\bibinfo {title}
  {The odd free surface flows of a colloidal chiral fluid},\ }\href@noop {}
  {\bibfield  {journal} {\bibinfo  {journal} {Nat. Phys.}\ }\textbf {\bibinfo
  {volume} {15}},\ \bibinfo {pages} {1188} (\bibinfo {year}
  {2019})}\BibitemShut {NoStop}%
\bibitem [{\citenamefont {Manikantan}\ and\ \citenamefont
  {Squires}(2020)}]{manikantan2020surfactant}%
  \BibitemOpen
  \bibfield  {author} {\bibinfo {author} {\bibfnamefont {H.}~\bibnamefont
  {Manikantan}}\ and\ \bibinfo {author} {\bibfnamefont {T.~M.}\ \bibnamefont
  {Squires}},\ }\bibfield  {title} {\bibinfo {title} {Surfactant dynamics:
  hidden variables controlling fluid flows},\ }\href@noop {} {\bibfield
  {journal} {\bibinfo  {journal} {J. Fluid Mech.}\ }\textbf {\bibinfo {volume}
  {892}},\ \bibinfo {pages} {P1} (\bibinfo {year} {2020})}\BibitemShut
  {NoStop}%
\bibitem [{\citenamefont {Stone}\ and\ \citenamefont
  {Ajdari}(1998)}]{stone1998}%
  \BibitemOpen
  \bibfield  {author} {\bibinfo {author} {\bibfnamefont {H.~A.}\ \bibnamefont
  {Stone}}\ and\ \bibinfo {author} {\bibfnamefont {A.}~\bibnamefont {Ajdari}},\
  }\bibfield  {title} {\bibinfo {title} {Hydrodynamics of particles embedded in
  a flat surfactant layer overlying a subphase of finite depth},\ }\href@noop
  {} {\bibfield  {journal} {\bibinfo  {journal} {J. Fluid Mech.}\ }\textbf
  {\bibinfo {volume} {369}},\ \bibinfo {pages} {151} (\bibinfo {year}
  {1998})}\BibitemShut {NoStop}%
\bibitem [{\citenamefont {Ramachandran}\ \emph {et~al.}(2011)\citenamefont
  {Ramachandran}, \citenamefont {Komura}, \citenamefont {Seki},\ and\
  \citenamefont {Gompper}}]{ramachandran2011}%
  \BibitemOpen
  \bibfield  {author} {\bibinfo {author} {\bibfnamefont {S.}~\bibnamefont
  {Ramachandran}}, \bibinfo {author} {\bibfnamefont {S.}~\bibnamefont
  {Komura}}, \bibinfo {author} {\bibfnamefont {K.}~\bibnamefont {Seki}},\ and\
  \bibinfo {author} {\bibfnamefont {G.}~\bibnamefont {Gompper}},\ }\bibfield
  {title} {\bibinfo {title} {Dynamics of a polymer chain confined in a
  membrane},\ }\href@noop {} {\bibfield  {journal} {\bibinfo  {journal} {Eur.
  Phys. J. E}\ }\textbf {\bibinfo {volume} {34}},\ \bibinfo {pages} {46}
  (\bibinfo {year} {2011})}\BibitemShut {NoStop}%
\bibitem [{\citenamefont {Oppenheimer}\ and\ \citenamefont
  {Diamant}(2010)}]{oppenheimer2010}%
  \BibitemOpen
  \bibfield  {author} {\bibinfo {author} {\bibfnamefont {N.}~\bibnamefont
  {Oppenheimer}}\ and\ \bibinfo {author} {\bibfnamefont {H.}~\bibnamefont
  {Diamant}},\ }\bibfield  {title} {\bibinfo {title} {Correlated dynamics of
  inclusions in a supported membrane},\ }\href@noop {} {\bibfield  {journal}
  {\bibinfo  {journal} {Phys. Rev. E}\ }\textbf {\bibinfo {volume} {82}},\
  \bibinfo {pages} {041912} (\bibinfo {year} {2010})}\BibitemShut {NoStop}%
\bibitem [{\citenamefont {Evans}\ and\ \citenamefont
  {Sackmann}(1988)}]{evans1988}%
  \BibitemOpen
  \bibfield  {author} {\bibinfo {author} {\bibfnamefont {E.}~\bibnamefont
  {Evans}}\ and\ \bibinfo {author} {\bibfnamefont {E.}~\bibnamefont
  {Sackmann}},\ }\bibfield  {title} {\bibinfo {title} {Translational and
  rotational drag coefficients for a disk moving in a liquid membrane
  associated with a rigid substrate},\ }\href@noop {} {\bibfield  {journal}
  {\bibinfo  {journal} {J. Fluid Mech.}\ }\textbf {\bibinfo {volume} {194}},\
  \bibinfo {pages} {553} (\bibinfo {year} {1988})}\BibitemShut {NoStop}%
\bibitem [{\citenamefont {Abramowitz}\ and\ \citenamefont
  {Stegun}(1972)}]{abramowitzhandbook}%
  \BibitemOpen
  \bibfield  {author} {\bibinfo {author} {\bibfnamefont {M.}~\bibnamefont
  {Abramowitz}}\ and\ \bibinfo {author} {\bibfnamefont {I.~A.}\ \bibnamefont
  {Stegun}},\ }\href@noop {} {\emph {\bibinfo {title} {Handbook of Mathematical
  Functions}}}\ (\bibinfo  {publisher} {Dover},\ \bibinfo {address} {New
  York},\ \bibinfo {year} {1972})\BibitemShut {NoStop}%
\bibitem [{\citenamefont {Saffman}(1976)}]{saffman1976}%
  \BibitemOpen
  \bibfield  {author} {\bibinfo {author} {\bibfnamefont {P.~G.}\ \bibnamefont
  {Saffman}},\ }\bibfield  {title} {\bibinfo {title} {Brownian motion in thin
  sheets of viscous fluid},\ }\href@noop {} {\bibfield  {journal} {\bibinfo
  {journal} {J. Fluid Mech.}\ }\textbf {\bibinfo {volume} {73}},\ \bibinfo
  {pages} {593} (\bibinfo {year} {1976})}\BibitemShut {NoStop}%
\bibitem [{\citenamefont {Ramachandran}\ \emph {et~al.}(2010)\citenamefont
  {Ramachandran}, \citenamefont {Komura}, \citenamefont {Imai},\ and\
  \citenamefont {Seki}}]{ramachandran2010}%
  \BibitemOpen
  \bibfield  {author} {\bibinfo {author} {\bibfnamefont {S.}~\bibnamefont
  {Ramachandran}}, \bibinfo {author} {\bibfnamefont {S.}~\bibnamefont
  {Komura}}, \bibinfo {author} {\bibfnamefont {M.}~\bibnamefont {Imai}},\ and\
  \bibinfo {author} {\bibfnamefont {K.}~\bibnamefont {Seki}},\ }\bibfield
  {title} {\bibinfo {title} {Drag coefficient of a liquid domain in a
  two-dimensional membrane},\ }\href@noop {} {\bibfield  {journal} {\bibinfo
  {journal} {Eur. Phys. J. E}\ }\textbf {\bibinfo {volume} {31}},\ \bibinfo
  {pages} {303} (\bibinfo {year} {2010})}\BibitemShut {NoStop}%
\bibitem [{\citenamefont {Strogatz}(1994)}]{strogatz}%
  \BibitemOpen
  \bibfield  {author} {\bibinfo {author} {\bibfnamefont {S.~H.}\ \bibnamefont
  {Strogatz}},\ }\href@noop {} {\emph {\bibinfo {title} {Nonlinear Dynamics and
  Chaos: With Applications to Physics, Biology, Chemistry, and Engineering}}}\
  (\bibinfo  {publisher} {Westview, Boulder},\ \bibinfo {year}
  {1994})\BibitemShut {NoStop}%
\bibitem [{\citenamefont {Hargus}\ \emph {et~al.}(2020)\citenamefont {Hargus},
  \citenamefont {Klymko}, \citenamefont {Epstein},\ and\ \citenamefont
  {Mandadapu}}]{hargus2020}%
  \BibitemOpen
  \bibfield  {author} {\bibinfo {author} {\bibfnamefont {C.}~\bibnamefont
  {Hargus}}, \bibinfo {author} {\bibfnamefont {K.}~\bibnamefont {Klymko}},
  \bibinfo {author} {\bibfnamefont {J.~M.}\ \bibnamefont {Epstein}},\ and\
  \bibinfo {author} {\bibfnamefont {K.~K.}\ \bibnamefont {Mandadapu}},\
  }\bibfield  {title} {\bibinfo {title} {Time reversal symmetry breaking and
  odd viscosity in active fluids: Green--kubo and nemd results},\ }\href@noop
  {} {\bibfield  {journal} {\bibinfo  {journal} {J. Chem. Phys.}\ }\textbf
  {\bibinfo {volume} {152}},\ \bibinfo {pages} {201102} (\bibinfo {year}
  {2020})}\BibitemShut {NoStop}%
\bibitem [{\citenamefont {Svetlizky}\ and\ \citenamefont
  {Roichman}(2021)}]{svetlizky2021}%
  \BibitemOpen
  \bibfield  {author} {\bibinfo {author} {\bibfnamefont {I.}~\bibnamefont
  {Svetlizky}}\ and\ \bibinfo {author} {\bibfnamefont {Y.}~\bibnamefont
  {Roichman}},\ }\bibfield  {title} {\bibinfo {title} {Spatial crossover
  between far-from-equilibrium and near-equilibrium dynamics in locally driven
  suspensions},\ }\href@noop {} {\bibfield  {journal} {\bibinfo  {journal}
  {Phys. Rev. Lett.}\ }\textbf {\bibinfo {volume} {127}},\ \bibinfo {pages}
  {038003} (\bibinfo {year} {2021})}\BibitemShut {NoStop}%
\bibitem [{\citenamefont {Tauber}\ \emph {et~al.}(2019)\citenamefont {Tauber},
  \citenamefont {Delplace},\ and\ \citenamefont {Venaille}}]{tauber2019}%
  \BibitemOpen
  \bibfield  {author} {\bibinfo {author} {\bibfnamefont {C.}~\bibnamefont
  {Tauber}}, \bibinfo {author} {\bibfnamefont {P.}~\bibnamefont {Delplace}},\
  and\ \bibinfo {author} {\bibfnamefont {A.}~\bibnamefont {Venaille}},\
  }\bibfield  {title} {\bibinfo {title} {A bulk-interface correspondence for
  equatorial waves},\ }\href@noop {} {\bibfield  {journal} {\bibinfo  {journal}
  {J. Fluid Mech.}\ }\textbf {\bibinfo {volume} {868}},\ \bibinfo {pages} {R2}
  (\bibinfo {year} {2019})}\BibitemShut {NoStop}%
\bibitem [{\citenamefont {Tauber}\ \emph {et~al.}(2020)\citenamefont {Tauber},
  \citenamefont {Delplace},\ and\ \citenamefont {Venaille}}]{tauber2020}%
  \BibitemOpen
  \bibfield  {author} {\bibinfo {author} {\bibfnamefont {C.}~\bibnamefont
  {Tauber}}, \bibinfo {author} {\bibfnamefont {P.}~\bibnamefont {Delplace}},\
  and\ \bibinfo {author} {\bibfnamefont {A.}~\bibnamefont {Venaille}},\
  }\bibfield  {title} {\bibinfo {title} {Anomalous bulk-edge correspondence in
  continuous media},\ }\href@noop {} {\bibfield  {journal} {\bibinfo  {journal}
  {Phys. Rev. Res.}\ }\textbf {\bibinfo {volume} {2}},\ \bibinfo {pages}
  {013147} (\bibinfo {year} {2020})}\BibitemShut {NoStop}%
\bibitem [{\citenamefont {Liebchen}\ \emph {et~al.}(2018)\citenamefont
  {Liebchen}, \citenamefont {Monderkamp}, \citenamefont {Ten~Hagen},\ and\
  \citenamefont {L{\"o}wen}}]{liebchen2018}%
  \BibitemOpen
  \bibfield  {author} {\bibinfo {author} {\bibfnamefont {B.}~\bibnamefont
  {Liebchen}}, \bibinfo {author} {\bibfnamefont {P.}~\bibnamefont
  {Monderkamp}}, \bibinfo {author} {\bibfnamefont {B.}~\bibnamefont
  {Ten~Hagen}},\ and\ \bibinfo {author} {\bibfnamefont {H.}~\bibnamefont
  {L{\"o}wen}},\ }\bibfield  {title} {\bibinfo {title} {Viscotaxis:
  Microswimmer navigation in viscosity gradients},\ }\href@noop {} {\bibfield
  {journal} {\bibinfo  {journal} {Phys. Rev. Lett.}\ }\textbf {\bibinfo
  {volume} {120}},\ \bibinfo {pages} {208002} (\bibinfo {year}
  {2018})}\BibitemShut {NoStop}%
\end{thebibliography}

%

\end{document}